\documentclass[12pt,letterpaper]{JHEP3}

\input{epsf}
\usepackage{epsfig}
\usepackage{cite}

\def\text{}

\def\pa{\partial}

\def\to{\rightarrow}
\def\be{\begin{equation}}
\def\ee{\end{equation}}
\def\bea{\begin{eqnarray}}
\def\eea{\end{eqnarray}}
\def\nonu{\nonumber \\{}}
\def\half{{1 \over 2}}
\def\ca{{\cal{A}}}
\def\cb{{\cal{B}}}
\def\cf{{\cal{F}}}
\def\cg{{\cal{G}}}

\def\cm{{\cal{M}}}
\def\cn{{\cal{N}}}

\def\sF{{{\rm F}\!\!\!\!\hskip.8pt\hbox{\raise1pt\hbox{/}}\,}}

\def\a{\alpha}

\def\d{\delta}
\def\e{\epsilon}

\def\f{\phi}

\def\m{\mu}
\def\n{\nu}
\def\o{\omega}
\def\p{\pi}

\def\r{\rho}
\def\s{\sigma}

\def\F{\Phi}
\def\G{\Gamma}

\def\L{\Lambda}
\def\O{\Omega}

\def\S{\Sigma}

\def\norm{\beta}

\newcommand\tfbox[1]{\setlength\fboxsep{0pt}\fbox{#1}}

\title{\begin{center} 5D fuzzball geometries and 4D polar states \end{center}}
\author{\begin{center}{Joris Raeymaekers, Walter Van Herck, Bert Vercnocke and Thomas Wyder}\end{center}\\
\begin{center}{Institute for Theoretical Physics}\\
{K.U. Leuven }\\
{Celestijnenlaan 200D}\\
{B-3001 Leuven, Belgium}\end{center}
\bigskip
\centerline{{\rm E-mail}: \email{joraey, waltervh, bert, thomas .at.itf.fys.kuleuven.be}}}

\abstract{We analyze the map between a class of `fuzzball' solutions
in five dimensions and four-dimensional multicentered solutions under the  4D-5D connection,
and interpret the resulting configurations in the framework of Denef and Moore \cite{Denef:2007vg}.
In five dimensions, we consider Kaluza-Klein monopole  supertubes with circular profile which
represent microstates of a small
black ring.  The resulting four-dimensional configurations are, in a suitable duality frame, polar states
consisting of stacks of D6 and anti-D6 branes with flux. We argue that these four-dimensional
configurations represent zero-entropy constituents of a 2-centered configuration where one of the
centers is a small black hole. We also discuss how spectral flow transformations in five dimensions,
leading to configurations with momentum,  give rise to four-dimensional  D6 anti-D6 polar configurations with
different flux distributions at the centers.}

\preprint{KUL-TF-08-09}

\begin{document}

\section{Introduction and summary}

Recent years have seen a significant progress in the understanding of the
supergravity description of BPS states of string theory, both in four and five noncompact
dimensions. In four dimensions, it has been established that BPS states of a given
charge are often realized as multicentered solutions in supergravity \cite{Denef:2000nb,Denef:2001xn,Bates:2003vx,Denef:2007vg,deBoer:2008fk}.
An important class of multicentered configurations are the `polar' states for which no single-centered solution exists and  which
contribute to the polar part of the OSV partition function \cite{Ooguri:2004zv} regarded as a
generalized modular form. From the knowledge of their microscopic degeneracies, the full partition function was  reconstructed in \cite{Denef:2007vg},
leading to a derivation of an OSV-type relation.
Another important type of configurations  are the so-called `scaling' solutions, which carry the same charges as a (large)
black hole and can be seen as a deconstruction of the black hole into zero-entropy constituents \cite{Denef:2007yt}.

On the five-dimensional side as well, the BPS objects are not restricted to single-centered
black holes. There also exist supersymmetric black rings and  black hole-black ring composites  \cite{Elvang:2004rt,Bena:2004de,Elvang:2004ds}, see  \cite{Emparan:2006mm} for review
and a more complete list of references.
There are also Kaluza-Klein monopole supertube
solutions which carry the charges of a black hole or black ring and are smooth and horizonless \cite{Maldacena:2000dr,Lunin:2001jy,Lunin:2002qf,Lunin:2002iz,Lunin:2004uu,Giusto:2004ip,Giusto:2004id,Giusto:2004kj,
Bena:2005ay,Taylor:2005db,Saxena:2005uk,Giusto:2005ag,Giusto:2006zi,Kanitscheider:2007wq}.
These can be seen as gravity duals to individual microstates in the CFT description of the black hole, leading to
the `fuzzball' picture proposed by Mathur and collaborators (see \cite{Mathur:2005zp,Skenderis:2008qn} for reviews and further references). In this proposal, the black hole horizon is an artefact of
 an averaging procedure over an ensemble of such smooth solutions.

These zoos of four and five-dimensional BPS configurations are not unrelated, and it is often possible to
continuously interpolate between 4D and 5D configurations using the `4D-5D connection' \cite{Gaiotto:2005gf,Gaiotto:2005xt,Elvang:2005sa,Bena:2005ni,Behrndt:2005he,Ford:2007th}.
Five-dimensional configurations can often be embedded in Taub-NUT space in a supersymmetric manner.
The spatial geometry of Taub-NUT space interpolates between ${\bf R}^4$ near the origin and ${\bf R}^3 \times S^1$ at infinity.
By varying the size of the $S^1$, one can then interpolate between effectively five and four-dimensional configurations.
Under this map, a point-like configuration at the center of Taub-NUT space becomes a 4D pointlike solution with
added Kaluza-Klein monopole charge. A ring-like configuration at some distance from the center goes over into a two-centered
solution where one center comes from the wrapped ring and the other contains Kaluza-Klein monopole charge. Angular momentum
 in 5D goes over into linear momentum along $S^1$ in four dimensions.

The goal of the current work is to give an explicit mapping between  supertube solutions arising in the fuzzball
picture in five dimensions and multi-centered solutions in four dimensions under the 4D-5D connection, and to
interpret the resulting configurations using the tools developed in \cite{Denef:2007vg}.
We will work in toroidally compactified  type II string theory, and consider a symmetric class of 2-charge  supertubes which are described
by a circular profile \cite{Maldacena:2000dr,Lunin:2001jy,Lunin:2002qf,Lunin:2002iz}, as well as 3-charge solutions obtained from those under spectral flow
\cite{Lunin:2004uu,Giusto:2004ip,Giusto:2004id,Giusto:2004kj}.
Placing such  supertubes  in Taub-NUT space gives the solutions that were constructed
in \cite{Bena:2005ay,Saxena:2005uk}. Applying the 4D-5D connection, we will show that,
in  the standard type IIB duality frame, one obtains  4D solutions which  are two-centered
 Kaluza-Klein monopole-antimonopole pairs  carrying flux-induced D1 and D5-brane charge and momentum.
 These solutions can be described within an STU-model truncation of $N=8$ supergravity
and  can be seen as simple examples of `bubbled' solutions \cite{Bena:2005va,Berglund:2005vb,Bena:2006is,Balasubramanian:2006gi,Bena:2006kb,Cheng:2006yq,Bena:2007ju,Gimon:2007mha,Bena:2007qc}
(for a review, see (\cite{Bena:2007kg}).
To make contact with the techniques developed for analyzing multicentered  configurations in Calabi-Yau
compactifications, we will
transform these configurations to a type IIA duality frame where all charges and dipole moments carried
arise from a D6-D4-D2-D0 brane system.
In this duality frame, the relevant configurations are two stacks of D6-branes and anti-D6 branes
with worldvolume fluxes turned on.
Those configurations fall into the class of `polar' states in 4D for which no single centered solution exists.

Let us briefly summarize our results. We consider 5D supertube solutions carrying D1 charge $N_1$, D5 charge $N_5$ and momentum $P$ and
which
are the gravity duals of a class of symmetric states in the D1-D5 CFT  with quantum numbers
\be \begin{array}{lcl}
L_0 = N_1 N_5 \left(m^2 + {m\over n} + 1/4\right)\,, & \qquad & \bar L_0 = {N_1 N_5\over 4}\,, \\
J^3 = - {N_1 N_5 \over 2} \left( 2 m + {1\over n} \right)\,, & \qquad & \bar J^3 = -{N_1 N_5 \over 2 n}\,,\\
P = L_0 - \bar L_0 = N_1 N_5 m \left(m+ {1 \over n}\right)\,. & &
\end{array} \label{CFTqn}
\ee
These represent Ramond sector states that are in a right-moving ground state and, on the left-hand side,
excited states in a twisted sector. The integer $n$ labels the twist sector and should
be a divisor of $N_1 N_5$. In a component string picture, $n$ represents the length of the component strings.
These states can be seen as obtained from Ramond ground states through a left-moving spectral
flow transformation  determined by the  parameter $m$, which should be an integer.
They carry momentum only when $m$ is nonzero.

After applying the 4D-5D connection to these configurations, we will interpret them in a U-dual
type IIA frame where all the charges arise from D6-D4-D2-D0 branes. Only 4 electric charges $q_I$ and   magnetic
charges $p^I$
are turned on  in these solutions. They arise from wrapping D-branes on the internal cycles
given in table \ref{chargesframeA}.

\TABLE{}{
\begin{table}\begin{center}
\begin{tabular}{cl|cl}
$q_0:$& $D0      $      & $p^0:$& $D6(T_1 \times T_2 \times T_3)$\\
$q_1:$& $D2(T_1) $      & $p^1:$& $D4(T_2 \times T_3)$\\
$q_2:$& $D2(T_2) $      & $p^2:$& $D4(T_1 \times T_3)$\\
$q_3:$& $D2(T_3) $      & $p^3:$& $D4(T_1 \times T_2)$\\
\end{tabular}\end{center}\caption{Type IIA D-brane charges carried by  our configurations. We have denoted  the submanifold wrapped by the
brane in brackets.}\label{chargesframeA}\end{table}}

Under the 4D-5D connection, the 5D quantum numbers (\ref{CFTqn}) map to the following 4D charges
\be
\begin{array}{cccccc}
5D:& N_1& N_5& J^3 & \bar J^3 & P \\
4D: & p^2 & p^3 & - {q_1\over 2}& - J_{z} & - q_0 \\
\end{array}\label{chargedict}\ee
Writing charges as $\G = (p^I , q_I)$, the 4D BPS state
corresponding to (\ref{CFTqn}) carries the charge \be \G_{\rm tot}
= \left( 0,1,N_1,N_5,\left(2m + {1\over n}\right)N_1 N_5,0,0,-
m\left(m+{1\over n}\right) N_1 N_5\right)\,.\ee This is a polar
charge   for which  there is no single-centered solution. It is
realized as a two-centered solution consisting of two stacks of D6
and anti-D6 branes with fluxes. Writing the charge as an element of the even cohomology
as  we will
explain in section \ref{Aframe}, the charges are \bea  \G_1 &=&-
n e^{-\left( m + {1\over n} \right) \o_1 + m N_1  \o_2 + m N_5
\o_3}\,,
 \nonu
 \G_2 &=&  n e^{-m  \o_1 + \left(  m + {1 \over
n}\right)N_1 \o_2 +  (m+ {1\over n}) N_5 \o_3} \,.
\label{intro4Dcharges}
\eea
The length of the component string $n$ has become the number of D6 and anti-D6 branes in the 4D picture, while
the spectral flow parameter $m$  has become a flux parameter. The restrictions on these parameters from charge quantization
match the quantization conditions in the CFT description.

This paper is organized as follows. In section \ref{Aframe}, we review the construction of
multicentered solutions in the STU-model and construct the solutions with charges (\ref{intro4Dcharges}).
We explain why these are polar states and review  the corresponding split attractor flow trees. In section \ref{Bframe},
we transform to a U-dual type IIB duality
 frame and discuss the lift of our solutions to 10 dimensions. We show that the solutions represent supertubes
embedded in Taub-NUT space,
and discuss the 5D limit. In section \ref{micro}, we discuss the microscopic interpretation of our configurations from the 4D and 5D points of view. We end with some prospects for future research in section \ref{concl}.
In appendix \ref{reductionBframe}, we discuss in detail the reduction formulae in the type IIB duality frame.

\section{A class of polar states in $N=8$ supergravity} \label{Aframe}
In this section we construct 2-center solutions in type IIA on a six-torus containing D6 and anti-D6 branes with flux (\ref{intro4Dcharges}),
and discuss the corresponding split attractor flow.
 These solutions  can be described in a truncation to an STU-model which we presently review.

\subsection{STU-truncation of type IIA on $T^6$}

We consider type IIA string theory compactified on a six-torus, which reduces in the low-energy limit to $N=8$ supergravity in four dimensions.
In  $N=2$ language, the $N=8$ gravity multiplet decomposes into the $N=2$  gravity multiplet, 6 gravitini multiplets, 15 vector multiplets, and 10 hypermultiplets.
For our purposes, it will be
sufficient to consider a
consistent truncation to a sector where only gravity and 3 vector multiplets are excited. This sector
is described by
the well-known STU model \cite{Duff:1995sm,Behrndt:1996hu} consisting of  $N=2$ supergravity coupled to 3 vector multiplets
with symmetric prepotential
$$ F = - D_{ABC} { X^A X^B X^C \over X^0} = - {X^1 X^2 X^3 \over X^0}\,, $$
where $D_{ABC}= {1 \over 6} |\e_{ABC}|$.
The bosonic part of the action is given by
\bea   S &=& {1\over 16 \p
G_4 } \int d^4 x \sqrt{-G}\Big[ R -\half \sum_{A=1}^3 {\pa_\m z^A
\pa^\m \bar z^A\over ({\rm Im} z^A)^2}\nonu
&& + {\norm^2 \over 2 } {\rm Im}
\cn_{IJ} \cf^I_{\m\n}\cf^{J\;\m\n}  + {\norm^2 \over 4 } {\rm Re}
\cn_{IJ}\e^{\m\n\r\s}
\cf^I_{\m\n}\cf^J_{\;\r\s}\Big]\label{STUaction} \,.\eea
with $z^A =
X^A/X^0 \equiv a_A + i b_A,\ A =1,2,3,\ I = 0,1,2,3$ and $\e^{0123
}\equiv 1$. We have left an arbitrary normalization constant $\norm$ in front of the
kinetic terms of the $U(1)$ fields for easy comparison with different conventions used in the
literature.
The matrix $\cn$ is given by
\be
\cn_{IJ} = \bar F_{IJ} + 2 {\rm i}{ {\rm Im} (F_{IK}) X^K{\rm Im} (F_{JL}) X^L \over {\rm Im} (F_{MN}) X^M X^N }\,.
\label{NfromF}
\ee
where $F_{IJ} = {\pa \over \pa X^I}{\pa \over \pa X^J} F $.
The explicit form of $\cn$ can be found in the Appendix (\ref{Ngen}). In our conventions,
the scalars $b_A$ have to be positive in order to have the correct kinetic term for the $U(1)$ fields.

We will, for simplicity, choose the hypermultiplet moduli such
that  the six-torus is metrically a product of three 2-tori $T_1
\times T_2 \times T_3$\footnote{For later convenience, we also
take $T_1$ to be rectangular and denote its two circles by $S^4 ,
S^5$.}. The 10-dimensional origin of the fields in
(\ref{STUaction}) is the following. The vector multiplet scalars
$z^A = X^A/X^0, A= 1,2,3$ describe complexified K\"ahler
deformations of the tori  $T_A$: \be B + i J = z^A \o_A\,, \ee
where $\o_A$  are normalized volume forms on $T_A$ satisfying
$\int_{T_A} \o_B = \d^A_B$. The constants $D_{ABC}$ entering in
the prepotential are proportional to the intersection numbers:
$D_{ABC} = {1 \over 6 } \int \o_A \wedge \o_B \wedge \o_C$. The
four $U(1)$ field strengths $\cf^I = d\ca^I, I = 0, \ldots , 3$
arise from dimensional reduction of the RR sector. Charged BPS
states  can carry electric and magnetic charges under the four
$U(1)$ fields. We will denote the magnetic charges by $p^I$ and
the electric charges by $q_I$ and write a general charge vector
$\G$ either by a row vector or an element of the even  cohomology
of $T^6$: \be \G = (p^0,p^A,q_A,q_0) = p^0 + p^A \o_A + q_A \o^A +
q_0 \o_{\rm vol}\,,\ee with $\o^A = 3 D_{ABC} \o_B \wedge \o_C$
and $ \o_{\rm vol} = \o_1 \wedge  \o_2 \wedge  \o_3$ and $A= 1,
\ldots, 3$. Taking into account charge quantization, the
components $p^I, q_I$ should be integers or $\G \in H^{\rm
even}(T^6, {\bf Z})$. We also define a symplectic inner product as
\be\langle \G ,\tilde \G \rangle = -p^0 \tilde q_0+ p^A \tilde
q_A-  q_A \tilde p^A + q_0 \tilde p^0 \,.\ee

{}From a 10-dimensional point of view, the charged BPS states arise
from D-branes wrapping  internal cycles. The D-brane
interpretation of the charges is  given in table
\ref{chargesframeA}. Dimensionally reducing the D-brane
Born-Infeld and Wess-Zumino action leads to point-particle source
terms to be added to the bulk action \cite{Billo:1999ip} (\ref{STUaction}): \be S_{\rm
source}  = { \norm \over  G_4} \int  \left[-   |Z(Q)| ds + {\norm \over 2 }
\langle Q, \ca \rangle \right]\,. \label{sourceaction}\ee Here,
$Q$ is a vector whose components have the dimension of length
 defined as
\be \int_{S^2} \cf^I = 4 \p Q^I \qquad \int_{S^2} \cg_I = 4 \p Q_I \ee
Where $\cg_I = {\rm Im}  \cn_{IJ} \star \cf^J + {\rm Re}  \cn_{IJ}
\cf^J$ and $\star$ denotes the Hodge dual.
For later convenience, as we will be taking the size of one of the internal directions to infinity,
 it will be useful  to work in conventions
 where we do not fix the coordinate volume of the internal cycles.
The components of $Q$ are then given by\footnote{To find agreement with
\cite{Bates:2003vx}, one should take  the coordinate volume of all cycles equal to one in units of $2\p \sqrt{\a'}$.
In that case, the relation between $Q$ and $\G$ is $Q = { \sqrt{G_4} \over \norm }\G$. Furthermore,
$\norm = 1$ in \cite{Bates:2003vx}.}
  \be Q^I = { \sqrt{8}  \over \norm}
T^I V^I G_4 p^I\,, \qquad Q_I = {\sqrt{8} \over \norm } T_I V_I G_4 q_I  \label{norm}\,.\ee
where $T^I, T_I$ are the tensions  of the  branes in table
\ref{chargesframeA}
and the $V^I, V_I$ are the coordinate volumes of the cycles they are wrapping. The quantity $Z(Q)$ in (\ref{sourceaction}) is the
central charge \be  Z = \langle Q, \O \rangle
\label{centralcharge}\,,\ee and $\O$ is the normalized period
vector defined as \be \O =  {\O_{\rm hol} \over \sqrt{8 b_1 b_2
b_3} }\,,\ee with $\O_{\rm hol}= - e^{z^A \o_A} $. A stack of
D-branes with worldvolume flux $F$ turned on sources lower D-brane
charges according to \be \G = {\rm Tr}\, e^F\,.\ee

We will denote this particular embedding of the STU model
in toroidally compactified type II string theory as `duality
frame A' in what follows. Later, in section \ref{Bframe}, we will also consider an embedding of the STU model into a
U-dual type IIB duality frame which we will call `frame B'.

\subsection{Multicentered BPS solutions}
We will now review the construction of general multicentered BPS
solutions in the STU model considered above, along the lines of
Bates and Denef \cite{Bates:2003vx}. Such  solutions can be constructed from
the harmonic functions \be  H^I = h^I + \sum_s { Q^I \over |\vec{x}
- \vec{x}_s| }; \qquad  H_I =  h_I + \sum_s { Q_I \over  |\vec{x} -
\vec{x}_s| }\,,\ee where the index $s$ runs over the centers and
$x_s$ are the locations of the centers in $\bf{R}^3$.

The metric and gauge fields are then completely determined
from the knowledge of a single  function $\S (H) $ on $\bf{R}^3$:
\be \S (H) = \sqrt{ {4 x_1 x_2 x_3 - L^2 \over (H^0)^2}}\,, \ee
with \bea x_A &=& 3 D_{ABC} H^B H^C - H_A H^0\,,\nonu L &=& 2 {H^1
H^2 H^3} + {  H_0}(H^0)^2  - {  H^A H_A H^0} \,.\eea If we replace
the harmonic functions $H$ in  $\S (H)$ by the charge vector $\G$,
the result is proportional to the Bekenstein-Hawking  entropy
$S(\G)$ of a black hole with charge vector $\G$: $\S (\G)=
S(\G)/\p$.

The
constants $h$ in the harmonic functions are related to the
asymptotic K\"ahler moduli as follows \be h = - {2 \over \norm}   {\rm Im} {\bar
Z_{\rm hol} \O \over |Z_{\rm hol} |} \ \Big{|}_\infty \,, \ee where
$Z_{\rm hol}$ is the holomorphic central charge \be  Z_{\rm hol} = \langle \G_{\rm
tot}, \O_{\rm hol} \rangle \,.\label{holcentralcharge}\ee Of the 8
components of $h$, only 6 are independent, corresponding to the
asymptotic values of the 6 moduli $a_A,\ b_A$. Indeed, from the
expressions above it follows that the $h$ satisfy two constraints
\bea \S ( h) &=& {1 \over \norm^2} \,,\nonu \langle h ,Q_{tot} \rangle &=& 0\,.
\label{hconstr} \eea

The metric of the multi-centered solution is given by \be
ds_4^2=-{ 1 \over\norm^2  \S(H) }(dt+  \omega )^2+{\norm^2 \S(H) } d \vec x^2
\label{4Dmetric}\,,\ee where $\omega$  is a 1-form on $\bf{R}^3$
that satisfies \be \star_3 d\omega = \norm^2 \langle dH, H\rangle =\norm^2 \left( -H_0
dH^0+H_A dH^A -H^A dH_A + H^0 dH_0\right) \label{omega}\,,\ee where the
Hodge star $\star_3$ is to be taken with respect to the flat
metric on $\bf{R}^3$. The integrability condition for the
existence of $\o$ leads to constraints on the positions  of the
centers: \be   \sum_t  {\langle Q_s ,Q_t \rangle \over |x_s - x_t|
}+ \langle Q_s , h \rangle = 0\,.\ee An important condition for
the existence of the supergravity solution is that, when the above
conditions are imposed, the function $\S(H)$ should be  real
everywhere. Multicenter solutions whose charges are non-parallel
also carry angular momentum given by
\be \vec{J} = \half \sum_{s
<t} \langle \G_s, \G_t \rangle {\vec{x}_s - \vec{x}_t \over
|\vec{x}_s - \vec{x}_t|}\, . \ee

In the special case of only 2 centers, the constraint on the distance $a$ between the centers simplifies
to
\be
 a = {\langle Q_1 ,Q_2 \rangle \over \langle Q_2, h \rangle }\,,\label{distance}
\ee
while the angular momentum is
\be
J_z = \half  \langle \G_1, \G_2 \rangle\,, \label{angmom}
\ee
where we have chosen the $z$-axis to run in the direction from the second to the first center.

The solution for the scalar moduli reads
\be z^A =  {  {\pa  \S(H)  \over \pa  H_A} - i H^A \over {\pa
\S(H) \over \pa  H_0}+ i H^0} \,.\ee More explicitly, splitting
$z^a$ into real and imaginary parts $z^A =  a_A + i b_A,\ A
=1,2,3$ one finds \bea a_A  &=& -{H^A \over H^0} + {L \over 2 x_A
H^0}\,\nonu b_A &=& {\S \over 2 x_A}\,. \eea

The gauge fields are given by \bea \ca^0 &=&  { 1\over \norm} {\pa \ln \S(H) \over
\pa H_0}({dt } +\omega)+ \ca^0_D\,,\nonu
 \ca^A &=&  - { 1\over \norm} {\pa \ln \S(H) \over \pa
H_A}({dt } +\omega)+ \ca^A_D\,, \eea where the Dirac parts
$\ca^I_D$ have to satisfy \be \star_3 d \ca^I_D =  d H^I
\label{diracparts}\,. \ee More explicitly,  one finds \bea \ca^0
&=& - { 1\over \norm} {L \over \S^2}  ({dt } + \o ) + \ca^0_D \,,\nonu
\ca^A &=& { 1\over \norm} { 6 D_{ABC} x_B x_C - H^A L \over H^0 \S^2}({dt } + \o ) + \ca^A_D \,.
\eea

These quantities can be worked out a little more explicitly as
\bea
\S &=&  \sqrt{ -4 H_0 H^1 H^2 H^3 -4  H^0 H_1 H_2 H_3 + (H^I H_I)^2 -2\sum_I (H_I)^2 (H^I)^2}\,,\label{Sigma}\\
a_A &=&
{  H_0 H^0  + H_A H^A -\sum_{B\neq A} H_B H^B\over
 6D_{ABC} H^B H^C - 2 H_A H^0}\,,\nonu
b_A &=&  {\S \over
6D_{ABC} H^B H^C - 2 H_A H^0}\,,\label{scalarsol}\\
\ca^0&=& {1 \over \norm  \S^2 }\left(H^0\left(  H_I H^I - 2 H_0 H^0
\right) - 2 D_{ABC} H^A H^B H^C  \right) ({dt } +
\o ) + \ca^0_D\,, \nonu
\ca^A &=& -{1 \over \norm  \S^2 }\left(H^A\left(
H_I H^I - 2 H^A H_A\right) - 6D_{ABC} H_B H_C H^0 \right) ({dt
} + \o ) + \ca^A_D\,. \label{U(1)s} \eea

We will also consider the effect of large gauge transformations of the $B$-field,  under which
 the $B$-field shifts with a harmonic form. Gauge invariance requires that this is accompanied by a
shift in the worldvolume flux, resulting in a transformation of the charge vector:
\be
B \to B + S\qquad \G \to e^S \G \,.\label{spectralflow}
\ee
In the 4D effective theory, the above transformation is induced by a symplectic transformation
\be
X^A \to X^A + S^A X^0.
\ee
Taking charge quantization into account, $S$ should be restricted
to be an element of the integer cohomology.
Large gauge  transformations
change the boundary conditions at infinity and, in the dual conformal field theory,
have the effect of inducing a spectral flow \cite{deBoer:2006vg,deBoer:2008fk,Bena:2008wt}.

\subsection{Solutions for polar states}
We will now describe a particular set of 2-centered solutions
where the centers are stacks of D6 and anti-D6 branes with
worldvolume fluxes turned on.  We will also show that for these
configurations  no
single-centered solutions with the same total charge exist. In the language of \cite{Denef:2007vg}, they  correspond to
polar states and are
described by attractor flow trees as we will review in paragraph \ref{flowtree}.

We will consider here two classes of  polar states: the first
class carries no D0-brane charge and has four net D4-D2 charges $p^1,
p^2,p^3,q_1$. These are the configurations (\ref{intro4Dcharges}) with $m=0$. By performing a  spectral flow transformation of the form
(\ref{spectralflow}) we will obtain a  second class of solutions ($m\neq 0$ in (\ref{intro4Dcharges})) which carry
the above four charges as well as D0-brane charge $q_0$.
In section \ref{Bframe} we
will show that these  two classes of configurations, after a U-duality
transformation,
give rise to smooth `fuzzball'
solutions placed in a Taub-NUT background.
The  solutions without D0-charge will map to fuzzball solutions with
D1-charge and D5-charge  in Taub-NUT space  while  the solutions carrying D0-charge
 will map to fuzzball solutions with D1-D5 charge and momentum P in Taub-NUT.

\subsubsection{Configurations without D0-charge}\label{solswoD0}
The first class of solutions we want to consider consists of a
stack of $n$ D6 branes and a stack of $n$ anti-D6 branes. Each
stack of branes has $U(n) = U(1) \times SU(n)$ gauge fields living
on the worldvolume. We will turn on worldvolume fluxes lying in
the $U(1)$ part so that each stack carries  lower-dimensional
D-brane charges as well. The fluxes we will turn on are characterized by
three numbers which, for later convenience, we will label $N_K, N_1, N_5$. The charges at the centers are \bea\label{twocenters}
 \G_1 &=& -n\, e^{-{N_K\over n}\o_1} = \left(-n,N_K,0,0,0,0,0,0\right)\,,\nonu
 \G_2 &=& n\, e^{{N_1 \over n} \o_2 + {N_5 \over n} \o_3 }=
\left(n,0,N_1,N_5,{N_1 N_5 \over n}, 0,0,0\right)\,.\label{chargesclass1} \eea
In the quantum theory,
charge quantization restricts $n, N_K, N_1, N_5$ to be integers
and   $n$ to be a divisor of $N_1 N_5$. These configurations carry
4 nonzero net charges $p^1, p^2,p^3,q_1$:
\be
\G_{\rm tot} = \left(0, N_K, N_1, N_5,{N_1 N_5\over n},0,0,0\right) \,.\label{totchargeclass1}
\ee
We will choose coordinates on ${\bf R}^3$ such that the first
center $\G_1$ is located at the origin and $\G_2$ lies on the
positive $z$-axis at $z=a$.
The harmonic functions are
\be\begin{array}{lll}
 H^0 = h^0 - {Q_n \over r} + {Q_n \over r_+ }\,,&\qquad  &  H_0 = h_0\,,\\
H^1 = h^1 + {Q_K  \over r }\,,& \qquad & H_1 = h_1 + {Q_1 Q_5  \over Q_n r_+ }\,, \\
H^2 = h^2 + {Q_1  \over r_+ }\,,& \qquad & H_2 = h_2\,, \\
H^3 = h^3 + {Q_5  \over r_+ }\,,& \qquad & H_3 = h_3\,. \\
\end{array}\label{harmfctnsclass1}
\ee We have defined $r_+$ to be the radial distance to the second
center: \be r_+ \equiv \sqrt{r^2 + a^2 - 2 a r \cos \theta}\,.\ee
From now on, we will choose the normalization constant $\norm$ in (\ref{STUaction}) to be
\be
\norm = {1 \over \sqrt 2}.
\ee
Using (\ref{norm}), the normalizations in the harmonic functions
are then  given by \be
\begin{array}{lcl}
Q_n = \half \sqrt{\a'} g n & \qquad &Q_K = { (2 \p)^2 (\a')^{3 \over 2} g \over 2 V_{T_1}}N_K\\
Q_1 = { (2 \p)^2 (\a')^{3 \over 2} g \over 2 V_{T_2}}N_1 & \qquad &Q_5 = { (2 \p)^2 (\a')^{3 \over 2} g \over 2 V_{T_3}}N_5
\end{array}
\label{normframeA}
\ee
where $g$ is the 10D  string coupling constant.

We can simplify the form of the solution by picking convenient values
for the asymptotic moduli and correspondingly the constants $h$.
We will choose six of the constants to be
\be
h_0 = -1 ;\qquad  \ h^1 = h^2 = h^3 =1 ; \qquad h_2 = h_3 = 0\,.\label{harmconst1a}
\ee
The remaining constants $h^0, h_1$ are then fixed by the constraints (\ref{hconstr}) to be
\be h_1 = - h^0 = {Q_1 Q_5 \over Q_n Q_K} \,.\label{harmconst1b}\ee
{}From (\ref{scalarsol}) we see that this choice of harmonic constants corresponds
to turning on asymptotic $B$-field on $T_1$ but not on $T_2,\ T_3$.

The constraint (\ref{distance}) on the distance between the
centers reads
\be a = {Q_K Q_1 Q_5  \over  Q_n^2-  Q_1 Q_5
  }\,.\label{posconstr}\label{distconstr1}\ee
The solution carries angular momentum given by (\ref{angmom}):
\be J_z = {N_K N_1 N_5 \over 2 n} . \label{angmomclass1}\ee

One can then find the explicit expressions for the metric, scalar fields and
$U(1)$ fields from (\ref{4Dmetric}, \ref{scalarsol},\ref{U(1)s}).  For configurations where $H_2 = H_3=0$, the expression (\ref{Sigma}) for $\S$ simplifies to
\be \S = \sqrt{ - 4 H_0 H^1 H^2 H^3 - ( H_0 H^0 - H_1 H^1)^2}\,.\ee

For the solution to the equations (\ref{omega}) and
(\ref{diracparts}) for $\o$ and the Dirac parts $\ca^I_D$ one
finds, using (\ref{posconstr}) and choosing convenient integration
constants,
\bea \o &=& {Q_K Q_1 Q_5  \over 2 a Q_n}
\left( {r + a \over r_+} -1 \right) ( \cos
\theta - 1 ) d\f\,, \nonu
\ca^0_D &=& Q_n\left(- \cos \theta
+ { r \cos \theta - a \over r_+}\right) d \f\,,\nonu
\ca^1_D &=& Q_K  \cos \theta  d \f\,,\nonu
\ca^2_D &=& Q_1  { r \cos \theta - a \over r_+} d
\f\,,\nonu
\ca^3_D &=& Q_5{ r \cos \theta - a \over r_+} d
\f \,.\label{omegasol} \eea

\subsubsection{Spectral flow and adding D0-charge }\label{spectflow}

The second class of solutions we will be interested in is obtained from the ones
considered above by a spectral flow transformation of the form (\ref{spectralflow})
$ \G \to e^S \G .$ We can choose $S$ such that the new configuration carries nonzero
$p^1, p^2,p^3,q_1$ charges as well as D0-charge $q_0$, while keeping $q_2$ and $q_3$ zero. There is a one-parameter
family of spectral flows $S$ which does the job and which we will label by a parameter
$m$:
\be S = -m N_K  \o_1 +m N_1  \o_2 + m N_5  \o_3\,.\ee When taking
charge quantization into account, the parameter $m$ could be
fractional but such that  $m$ is a common multiple of $1/N_1,
1/N_5 $ and $1/N_K$. The charges carried by the two centers are
then the ones anticipated in (\ref{intro4Dcharges}) in the
introduction: \bea \G_1 &=&- n e^{-\left( m + {1\over n}
\right)N_K \o_1 + m N_1 \o_2 + m N_5  \o_3} \,,\nonu \G_2 &=&  n
e^{-m N_K \o_1 + \left(  m + {1 \over n}\right)N_1 \o_2 +  (m+
{1\over n}) N_5 \o_3}\,,\label{chargesclass2}
\eea
and the total charge of the solution is
\be
\G_{\rm tot} = \left( 0,N_K,N_1,N_5,\left(2m + {1\over n}\right)N_1 N_5,0,0,-
m\left(m+{1\over n}\right)N_K N_1 N_5\right)\,. \label{totchargeclass2}\ee
The angular momentum  of these configurations is independent of the parameter $m$ and still given by (\ref{angmomclass1}).
For $m=0$ we recover the configurations discussed in the previous section.

The harmonic functions for this configuration are
\be\begin{array}{lll}
 H^0 = h^0 - {Q_n \over r} + {Q_n \over r_+ }\,,&  &  H_0 = h_0 + {(mn + 1)(mn)^2 Q_K Q_1 Q_5 \over Q_n^2 r}
 - {(mn + 1)^2 mn Q_K Q_1 Q_5 \over Q_n^2 r_+}\,,\\
H^1 = h^1 + {(mn +1)Q_K  \over r }- {mn Q_K  \over r_+ }\,, &  & H_1 = h_1
- {(mn)^2 Q_1 Q_5  \over Q_n r} + {(mn + 1)^2 Q_1 Q_5  \over Q_n r_+ }\,, \\
H^2 = h^2 - {mnQ_1  \over r } + {(mn +1)Q_1  \over r_+ }\,,&  & H_2 = h_2
+  {(mn +1)mnQ_K Q_5  \over Q_n r } - {(mn +1)mnQ_K Q_5  \over Q_n r_+ }\,, \\
H^3 = h^3 - {mnQ_5  \over r } + {(mn +1)Q_5  \over r_+ }\,,&  & H_3 = h_3 +  {(mn +1)mn Q_K Q_1  \over Q_n r } -  {(mn +1)mn Q_K Q_1  \over Q_n r_+ }\,.\\
\end{array}\label{harmfctnsclass2}
\ee
As before, we  choose the asymptotic moduli such that $
h_0 = -1 ,\  \ h^1 = h^2 = h^3 =1 ,  \  h_2 = h_3 = 0$. The remaining constants
are determined by  (\ref{hconstr}) to be
\be h_1 = - h^0 = {(2 mn + 1) Q_1 Q_5 Q_n \over (mn + 1)mn Q_KQ_1Q_5 +Q_K Q_n^2} \,.\label{harmconst2b}\ee
For the constraint (\ref{distance}) on the distance one finds a rather complicated expression
{\footnotesize
\bea
{1 \over a}  &=& {1 \over  Q_K Q_1 Q_5 \left( (mn + 1)^2 (mn)^2 Q_1 Q_5 + Q_n^2\right)}  \Big{(}
  Q_n^4- Q_n^2\left( Q_1 Q_5 + (mn +1)(2 Q_1 Q_5 - Q_K(Q_1 + Q_5))\right)\nonu
& &+ (mn + 1)^2 (mn)^2 Q_1 Q_5 ( Q_K Q_1 + Q_K Q_5 + Q_1 Q_5)\Big{)}\,.\label{distconstrclass2}
\eea}

\subsection{Polarity and flow trees}\label{flowtree}
We will now describe how our configurations fit within the zoo of four-dimensional multicenter BPS solutions, using the tools
that were developed in \cite{Denef:2007vg}. Some well-founded conjectures put forth  there will allow us to draw conclusions which are valid beyond the leading supergravity approximation.
 We will now review some relevant points from \cite{Denef:2007vg}  to which we refer the reader for
more details.

The configurations we are considering here correspond to four-dimensional `polar' states.  Mathematically, polar states can be seen as the constituents of the polar part of the black hole partition function as a
generalized modular form. The full partition function can be reconstructed from the knowledge of the degeneracies of the polar
states, which was at the core of deriving an OSV-type relation for D4-D2-D0 black holes
in \cite{Denef:2007vg}.

Physically, the fact that a configuration is polar means that  no
single-centered solutions with these charges exist. For polar configurations, one can show that the
attractor flow equations that describe the radial evolution of the moduli fields always `crash'
at a regular\footnote{Regular meaning that the K\"ahler form on the internal space lies  within the K\"ahler cone.}  point in moduli space beyond which they cannot
be continued. This means  that a single-centered black hole solution cannot exist
in the supergravity approximation. Furthermore, by appropriately  choosing the asymptotic moduli,
one can show that, at the point where the attractor flow crashes, all curvatures remain small,
and hence this conclusion should not be modified by higher-derivative corrections to supergravity \cite{Moore:1998pn}.

As discussed in \cite{Denef:2007vg} the relevant quantity for
establishing whether a total charge system $\Gamma_{\rm{tot}}$ is
polar is the `reduced' D0 brane charge
\be\hat{q}_0=q_0-\frac{1}{2}D^{AB}q_A q_B\,,\ee where $D^{AB}=(6
D_{ABC}p^C)^{-1}$. If $\hat{q}_0>0$, the states are  polar and no
single centered black hole solutions carrying these charges exist.
For our configurations without D0-charge (\ref{totchargeclass1})
one obtains\be D^{AB} = \frac{1}{2 N_KN_1 N_5}\left( \matrix{
-N_K^2&N_KN_1&N_KN_5\cr N_KN_1&-N_1^2&N_1N_5\cr
N_KN_5&N_1N_5&-N_5^2 }\right),\ee and therefore \be
\hat{q}_0=\frac{N_KN_1N_5}{4n^2} .\label{qhat}\ee
 This means that these states are polar if we choose positive fluxes on our branes.
For $n=1$, when there is only
one D6 and one anti-D6 brane, $\hat{q}_0$
reaches its maximal value for given $p^1, p^2, p^3$ charge.
The quantity $\hat{q}_0$ is invariant under spectral flow transformations (\ref{spectralflow}), hence our
charge configurations with D0 charge (\ref{totchargeclass2}) are also polar with $\hat{q}_0$ still given by
(\ref{qhat}).

Even if  no single-centered solution exists, there can still be a  BPS state carrying
the desired charges which is realized as a multicentered configuration\footnote{Note that constituents need not be
`regular' black hole solutions, but can also be realized as `empty' holes where the center has zero entropy.}.
A proposed criterion to verify
whether such a BPS state exists
is whether there is an `attractor flow tree' for the given charge. This proposal is called the `split attractor
flow conjecture' and has been argued
 to establish the existence of the BPS state beyond the supergravity approximation.
An attractor flow tree is a graph in moduli space which starts at the background value of the moduli and
follows the single center
attractor flow until it hits a wall of marginal stability where it becomes energetically possible for the total charge to split into   two
constituents. There the flow splits in two parts corresponding to the single centered flows of the constituents.
This process is repeated until  one ends up at the attractor points for all the centers of the configuration.

We therefore now inspect the existence of  flow trees for our
charge configurations in order to be able to infer the existence
of the corresponding BPS state. We will show that the single
centered flow reaches a wall of marginal stability at a point
$z_{\rm split}$ in moduli space before reaching the crash point
$z_0$, where the single centered flow ends. At the marginal
stability wall, the flow branches into two flows representing the
D6 and anti D6 centers which reach their attractor points without
encountering any more marginal stability walls. A schematic
depiction of the split flow is given in figure \ref{flowtreefig}.

\FIGURE{}{
\begin{figure}[h]
\setlength{\unitlength}{1cm} 
\centering 
\begin{picture}(7,5.8) 
\put(0,0) {\tfbox{\includegraphics[width=0.5\textwidth,angle=0]{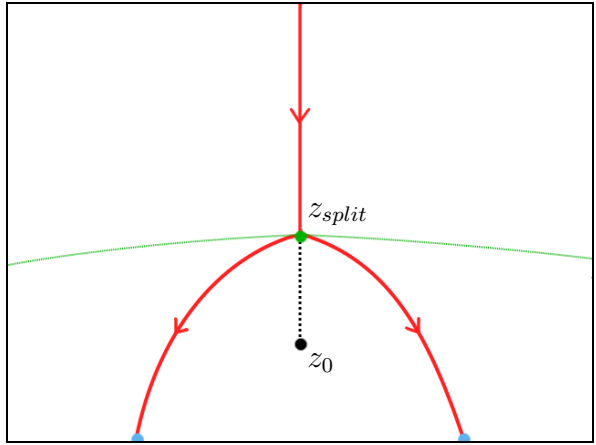}}}
\put(4,3) {$z_{split}$}
\put(4,1) {$z_0$}
\end{picture}
\caption{\label{flowtreefig} Schematic drawing of the split flow tree for our representative charge system. The flow coming in from the top (red line) reaches the wall of marginal stability (green line) at the splitpoint $z_{split}$ (green) before it would reach the crash point $z_0$ (black). One also sees the single flows for each center starting from the split point until they reach the boundary of moduli space (blue).}
\end{figure}}

A crucial simplification  is that, doing the spectral flow
transformation (\ref{spectralflow}), we can equivalently  examine the flow tree for a charge $e^S \G$ at
a shifted $B$-value $B + S$.
When, by shifting the asymptotic value of the $B$-field, one does not cross any walls of marginal stability,
we can simply fix the asymptotic $B$-field to a convenient value and choose a charge vector $e^S \G$ such
that the analysis becomes simple. This will be possible for our configurations, provided that we choose the background K\"ahler moduli large enough. The reason for this is
that walls of marginal stability between two charges can only run all the way to infinity for a `core-halo pair' of
D-branes (Halos can only carry D2-D0 brane charge, any other charge configuration will automatically be a core, see \cite{Denef:2007vg} for definitions and a more in-depth treatment of these concepts). Here, we are luckily always dealing with core constituents. From now on, we will take
the  asymptotic $B$-field to be zero and choose an appropriate charge vector $e^S \G$.

We can pick a charge representative by giving some convenient value to the spectral flow parameter $m$ in our
 general charge configuration (\ref{totchargeclass2}). We will take it
to have the value\footnote{As the
flow tree analysis takes place
within supergravity, we can ignore charge quantization restrictions for the moment.} $m = -{1 \over 2n}$.
This leads us to the total charge
\be \G_{\rm tot}=\left(0,N_K,N_1,N_5,0,0,0,\frac{N_KN_1N_5}{4n^2}\right)\,. \ee
This obviously is a pure D4-D0 system. As discussed above, we choose our background modulus to have purely imaginary and very large values, $z_{\infty}=(iy_{\infty}^1,iy_{\infty}^2,iy_{\infty}^3)$. The single centered flow
runs along the imaginary $z$-axes until the crash point is reached where the holomorphic central charge
(\ref{holcentralcharge}) vanishes. This happens at the point
\be z_0=i\sqrt{\frac{2\hat{q}_0}{6N_KN_1N_5}}(N_K,N_1,N_5)=i\frac{1}{\sqrt{12n^2}}(N_K,N_1,N_5) \,.\ee

Next one can check whether the flow hits a wall of marginal stability. This is per definition the locus where the phases of the central charges of the two centers align.
The charges at the centers read
\bea \G_1&=& \left(-n,\frac{N_K}{2},\frac{N_1}{2},\frac{N_5}{2},-\frac{N_1N_5}{4n},-\frac{N_KN_5}{4n},-\frac{N_KN_1}{4n},\frac{N_KN_1N_5}{8n^2}\right),\nonumber\\
\G_2&=&\left(n,\frac{N_K}{2},\frac{N_1}{2},\frac{N_5}{2},\frac{N_1N_5}{4n},\frac{N_KN_5}{4n},\frac{N_KN_1}{4n},\frac{N_KN_1N_5}{8n^2}\right)\,.
\eea
One easily sees that the real parts of the central charges are equal, whereas the imaginary parts have opposite signs. Thus, the wall will be hit when $\rm{Im}(Z_1) =\rm{Im}(Z_2)=0$. One finds \be z_{\rm split}=i\sqrt{\frac{3}{4n^2}}(N_K,N_1,N_5)\,. \ee As $\sqrt{\frac{3}{4n^2}}>\sqrt{\frac{1}{12n^2}}$ this means that the wall of marginal stability is always reached before the single flow crashes.
The single centered flows for the  fluxed D6 brane centers terminate at the boundary of moduli space
in the supergravity approximation.
Nevertheless they correspond to states in the BPS spectrum of string theory and higher derivative corrections
are expected to yield regular attractor points.

A further simple check also shows that the necessary stability criterion \cite{Denef:2002ru,Denef:2007vg} \linebreak $\langle\G_1,\G_2\rangle \cdot (\rm{arg}(Z_1)-\rm{arg}(Z_2)) > 0$ is met. This shows
that one indeed reaches the wall from the side where the single brane is stable and crosses to the side where the brane decays into a bound state. The condition
can be interpreted as ensuring that tachyonic strings would be present between the two constituent branes  on the `stable' side, in this case above the wall, such that a bound state is formed after tachyon condensation.

\section{U-duality and fuzzballs in Taub-NUT}\label{Bframe}
In this section, we would like to make contact between the polar solutions
constructed above and  various horizonless supertube solutions in five noncompact dimensions that are central to the fuzzball proposal advocated by Mathur and collaborators.
As a first step, we will make a duality transformation to a type IIB frame such that the charges
and dipole moments carried by our solutions are the same as the ones carried by the supertubes.

Let us briefly review these configurations. Fuzzball solutions in five noncompact dimensions can be seen
as Kaluza-Klein (KK) monopole\footnote{Recall that a Kaluza-Klein monopole in 10D is a 5+1-dimensional object whose transverse 4-dimensional space has Taub-NUT geometry or,
in the case of several centers, a Gibbons-Hawking space.}
supertubes where the KK monopole
charge is sourced along a contractible curve in 4  noncompact  directions.
One of the compact directions, which will become $S^4$ in our
conventions  (recall that we had denoted $T_1 = S^4 \times S^5$), is
a Taub-NUT circle which pinches off at every point of the curve. By adding flux to the KK-monopole, one can source the charge of D1 and D5-branes
wrapped around the $S^4$ circle. For a circular curve, one can place this configuration in a Taub-NUT space with a different Taub-NUT circle, $S^5$ in our
conventions, and interpolate between five and four  dimensions by varying the size of $S^5$.
We will show that the four-dimensional configurations obtained in this manner are U-dual to the D6-anti D6 polar solutions we discussed above.
\subsection{U-duality to a type IIB frame}
Let us first describe a U-duality transformation to a type IIB frame such that STU-model solutions lift to
configurations carrying the charges described above.
We will go to a duality frame
where  $p^0$ becomes a Kaluza-Klein monopole charge with Taub-NUT circle $S^4$, $p^1$ becomes a Kaluza-Klein monopole charge with
Taub-NUT circle $S^5$, $p^2$ becomes the charge of a $D1$-brane wrapped on $S^4$ and $p^3$ becomes
the charge of a $D5$-brane wrapped on $S^4\times T_2 \times T_3$. This is accomplished by making a
 U-duality transformation consisting of a T-duality along $S^4$, followed by  S-duality and 4
 T-dualities  along $T_1\times T_3$, as illustrated in table \ref{Udual}.
\TABLE{}{
\begin{table}
\begin{center}
\begin{tabular}{lclclcl}
IIA (frame A)& & IIB &  & IIB & &IIB (frame B)\\
D6 ($T^6$) & & D5 & &   NS5 & &  KK5 ($S^5 \times T_2 \times T_3$)\\
D4 ($T_2 \times T_3$) & T ($S^4$) & D5& S &  NS5&T ($S^4,S^5,T_3$)&  KK5 ($S^4 \times T_2\times T_3$)\\
D4 ($T_1 \times T_3 $) & $\longrightarrow$ &  D3& $\longrightarrow$ & D3&$\longrightarrow$&  D1 ($S^4$)\\
D4 ($T_1 \times T_2$) & &  D3 & &  D3& & D5 ($S^4\times T_2 \times T_3$)
\end{tabular}
\end{center}\caption{U-duality transformation from frame A to frame B}\label{Udual}\end{table}
}

This new duality frame will be denoted `frame B'. In this frame, the vector multiplet
scalars $z^1, z^2, z^3$ represent
the complex structure modulus of $T_1$, the 4D axion-dilaton and the (complexified)
K\"{a}hler modulus of $T_1$ respectively.
The $U(1)$ fields $\ca^0$ and $\ca^1$
are Kaluza-Klein gauge fields from the metric components $g_{\m 4}$ and $g_{\m 5}$ respectively, while
$\ca^2$ and $\ca^3$ arise from the RR two form components $C_{\m 4}$ and $C_{\m 5}$.
The 10-dimensional origin  of the full set of charges  in this frame is given in table \ref{frameBcharges}.
\TABLE{}{\begin{table}
\begin{center}
\begin{tabular}{cl|cl}
$q_0$& P ($S^4$)        & $p^0$& KK5 ($S^5\times T_2\times T_3$)\\
$q_1$& P ($S^5$) & $p^1$& KK5 ($S^4\times T_2\times T_3$)\\
$q_2$& D5 ($S^5\times T_2 \times T_3$)       & $p^2$& D1 ($S^4$)\\
$q_3$& D1 ($S^5$)& $p^3$& D5 ($S^4\times T_2 \times T_3$)\\
\end{tabular}\end{center}\caption{10D origin  of the charges in frame B}\label{frameBcharges}\end{table}}

In frame B, our first class of  polar solutions with charges (\ref{chargesclass1})
 corresponds to  two stacks of $n$ KK
monopoles and anti-KK monopoles with Taub-NUT circle $S^4$ carrying flux-induced charges of D1, D5, momentum and KK monopoles wrapped on
the $S^4$ circle. The more general
solutions (\ref{chargesclass2}) obtained by spectral flow carry momentum along $S^4$ as well.  Such
solutions will be smooth, and, as we will show, have the interpretation of KK monopole supertubes
embedded in Taub-NUT space.

\subsection{Lifting general multicenter solutions}
In order to see what our solutions look like in frame B from the 10-dimensional point of view, we need
to know the reduction formulas of type IIB on a six-torus to the four-dimensional STU-model action
(\ref{STUaction}) such that the 4D charges have the interpretation  given in table  \ref{frameBcharges}. This is worked out in detail
in  appendix \ref{reductionBframe}.

The metric of a general 4D multicentered solution lifts to a 10D
geometry where the $T_1$ torus is nontrivially fibered over the 4D
base: \bea ds^2_{10} &=& {1 \over \sqrt{ b^2  b^3} }ds^2_4 +
\sqrt{ b^2   b^3} \cm_{mn} (dx^m + \ca^{m-4})(dx^n +
 \ca^{n-4}) + \sqrt{ b^2 \over  b^3}ds^2_{T_2 \times
T_3}\,,\nonu ds_4^2&=&-{ 2 \over  \S }(dt+  \omega )^2+{ \S \over 2 } d \vec
x^2\,,\nonu \cm_{mn}  &=&  {1 \over  b^1}\left( \matrix{ ( a^1)^2
+( b^1)^2 & -a^1 \cr -a^1 &  1} \right), \qquad m,n=4,5\,. \eea The dilaton and RR
two-form are given by \bea e^{2 \F^{(10)}} &=&  {b^2 \over
b^3}\,,\nonu
C^{(10)} &=&  \half C_{\m\n} dx^\m dx^\n +  a^3 (dx^4
- \ca^{0})\wedge (dx^5 - \ca^{1}) \nonu && -
 dx^4 \wedge  \cb^2  -  dx^5 \wedge  \ca^3 +{1 \over 2 }
(\ca^0\wedge   \cb^2 + \ca^1 \wedge
 \ca^3)\,,\nonu
 d  a^2 &=& -(b^2)^2 \star F \,,\nonu
F &=& d C + {1 \over 4} (  \ca^0 \wedge  \cg^2 +   \cb^2 \wedge
\cf^0 + \ca^1 \wedge  \cf^3 +  \ca^3 \wedge  \cf^1)
\,.\label{10Dlift} \eea where the Hodge $\star$ is to be taken
with respect to the 4D metric $ds^2_4$.

It will be useful to rewrite the metric in the form of a lifted
solution of 6D supergravity as in \cite{Gutowski:2003rg,Bena:2004de,Bena:2008wt}, where the 6D part
of the metric is  written as a fibration over a 4D Gibbons-Hawking base space.
If both $p^0$ and $p^1$ are nonzero, both
the $S^4$ and $S^5$ are nontrivially fibered, and
we can choose either circle to be the fibre in the Gibbons-Hawking
geometry. Here, we will choose the  $S^5$ to be this fibre, so that
the  Gibbons-Hawking base space is
spanned by the coordinates $(r,\theta,\f,x^5)$. The metric can be rewritten in the form
\bea
ds^2 &=& -{ 1 \over H F} (dt + k)^2 +  { F \over H } \left( dx^4 - s - {1 \over F}(dt + k)\right)^2
+ H ds^2_{\rm GH} + \sqrt{ x_2 \over x_3} ds^2_{T_2 \times T_3}\,,\nonu
ds^2_{\rm GH} &=& {1 \over  H^1} (dx^5 +  \ca^{1}_D)^2 +  H^1 d{\bf x}^2\,.\label{6Dform}\eea
where we have defined
\bea
F &=&  \left( {H_2 H_3\over H^1}- H_0  \right)\,,\nonu
H &=& {\sqrt{  x_2 x_3} \over H_1}\,,\nonu
k &=&  \o +
{L H^1 - 2 x_2 x_3  \over2 H^0 (H^1)^2} (dx^5 +  \ca^{1}_D)\nonu
&=&  \o + {1 \over 2  H_1}\left( H_I H^I - 2 H_1 H^1 - {2 H_0 H_2 H_3\over H^1}\right)
(dx^5 +  \ca^{1}_D)\,,\nonu
s &=& - \ca^{0}_D+{H^0\over H^1}(dx^5 + \ca^{1}_D)\,.
\eea
We will now use these expressions to find the lift of our four-dimensional polar configurations.

\subsection{Lift of polar states without D0 charge}
We will first discuss the lift of our configurations (\ref{chargesclass1}) that do not
contain D0 charge in  frame A. In frame B   these correspond,  according to table \ref{frameBcharges},
 to  two stacks of $n$ KK
monopoles and anti-KK monopoles with Taub-NUT circle $S^4$ which carry flux-induced  D1, D5 and KK monopole
charges    wrapped on
the $S^4$ circle. We will now show that, from a 10D point of view, these charges precisely correspond to the Kaluza-Klein monopole supertubes
in Taub-NUT space that were constructed by Bena and Kraus in \cite{Bena:2005ay}.

The harmonic functions of the solution are given by (\ref{harmfctnsclass1}, \ref{harmconst1a}, \ref{harmconst1b}), where
the normalizations in the current duality frame should be taken to be, according to (\ref{norm}),
\be
\begin{array}{lcl}
Q_n ={n R_4\over 2}\,,& \qquad&Q_K = {N_K R_5\over 2}\,,\\
Q_1 =  {(2 \p)^4 g \a'^3 \over 2 R_5 V_{T_2\times T_3} }N_,\,.& \qquad &Q_5 = {g \a' \over 2 R_5} N_5\,.
\end{array}
\label{normframeB}
\ee
The constraint on the distance between the centers (\ref{distconstr1}) can also be written as
\be Q_n = \sqrt{Q_1 Q_5 \tilde H^1 }\,, \label{distconstrB}\ee
with $\tilde H^1 = 1 + {Q_K \over a}$.

We find  the lift of this class
of solutions to 10 dimensions in  duality frame B by plugging these expressions into (\ref{6Dform}).
Making a coordinate transformation
$x^4 \to x^4  + t$,
the metric becomes
\bea
ds^2 &=& { 1 \over \sqrt { H^2 H^3} } \left[ - (dt + k)^2  + ( dx^4 - s -  k)^2\right]
+ \sqrt { H^2 H^3}  ds^2_{\rm TN} + \sqrt{ H^2 \over H^3} ds^2_{T_2 \times T_3}\,,\nonu
ds^2_{\rm TN} &=& {1 \over H^1} (R_5 d \psi  + Q_K  \cos \theta  d \f)^2 +  H^1 d{\bf x}^2\,.
\label{6Dformclass1}
\eea
where we have defined the angle $\psi$ as $x^5 = R_5 \psi$.
{}From the ten-dimensional point of view,
the  constraint (\ref{distconstrB}) on the distance between the centers arises from requiring smoothness of the metric \cite{Bena:2005ay},
while the condition that $\S$ is real implies the absence of closed timelike curves \cite{Cheng:2006yq}.

The one-forms $k$ and $s$ have components along
$\f$ and $\psi$ and, using the distance constraint (\ref{distconstrB}), can be written as
\be
\begin{array}{lcl}
k_\psi = {R_5 Q_n  Q_K \over 2 a r r_+ \tilde H^1 H^1} \left[ r_+ - r - a - {2 a r \over Q_K}
\right]\,, & \qquad &
k_\f = {Q_n  Q_K \over 2 a  r_+ \tilde H^1 } \left[ r_+ - r - a + { r - a - r_+ \over H^1}
\cos \theta\right]\,, \\
s_\psi = { R_5 Q_n \over r  r_+  H^1 } \left[ r - r_+ - { r r _+ \over Q_K \tilde H^1}
\right]\,, & \qquad &
s_\f = {Q_n   \over   r_+  }\left[ a + { r_+ - r - {r_+ \over \tilde H^1} \over H^1}\cos \theta
\right]\, . \\
\end{array}\label{metricclassI}
\ee

Using (\ref{10Dlift})  one can show that the dilaton and RR three-form take the form
\bea
e^{2 \F} &=& {H^2 \over H^3}\,,\nonu
F^{(3)} &=& d \left[ { 1\over  H^2 } (dt + k) \wedge ( dx^4 -s-k )\right] - \star_4 d ( H^3)\,,
\label{3formclassI}
\eea
where the Hodge star $\star_4$ is to be taken with respect to the Taub-NUT metric $ds^2_{TN}$.

As we have argued, the above solutions represent the lift of a two-centered
KK-monopole anti-monopole system in frame B (or a D6 anti-D6 system in frame A),
where the Taub-NUT circle for these KK monopoles is the $S^4$. The KK monopoles sit
at a radial distance $r_+$ while the anti-monopoles sit at the origin. At the position of these centers,
the $S^4$ circle should pinch off. This is not so obvious in the 10D form of the metric (\ref{6Dformclass1}), so
let us illustrate this point in more detail here.
The coefficient in front of the $(dx^4)^2$ term in the metric (\ref{6Dformclass1})
is  $1 / \sqrt{ H^2 H^3}  $.  This factor goes to zero at $r = r_+$ but stays finite at $r=0$, so
it is not obvious that there is a KK anti-monopole source at the origin. Nevertheless, there should be such a source  since
the total KK monopole charge has to balance out, and it should be located at the origin because of symmetry reasons. The resolution to this puzzle lies in the fact
that the six-dimensional metric still contains a factor of  the six-dimensional
dilaton ${\rm e}^{\Phi^{(6)}}$. This factor is given by ${\rm e}^{\Phi^{(6)}}=  \frac 1 {b_2 b_3}$,
and hence the factor that measures the size of the $S^4$ is $b_2 b_3 / \sqrt{ H^2 H^3}$. One can easily
check that this factor indeed goes to zero both in $r=0$ and $r=r_+$. This is illustrated in figure \ref{pinchpict}.

\FIGURE{}{
\begin{figure}[h]
\setlength{\unitlength}{1cm}
\centering
\begin{picture}(15,5.8)
\put(0,0) {\tfbox{\includegraphics[width=0.45\textwidth,angle=0]{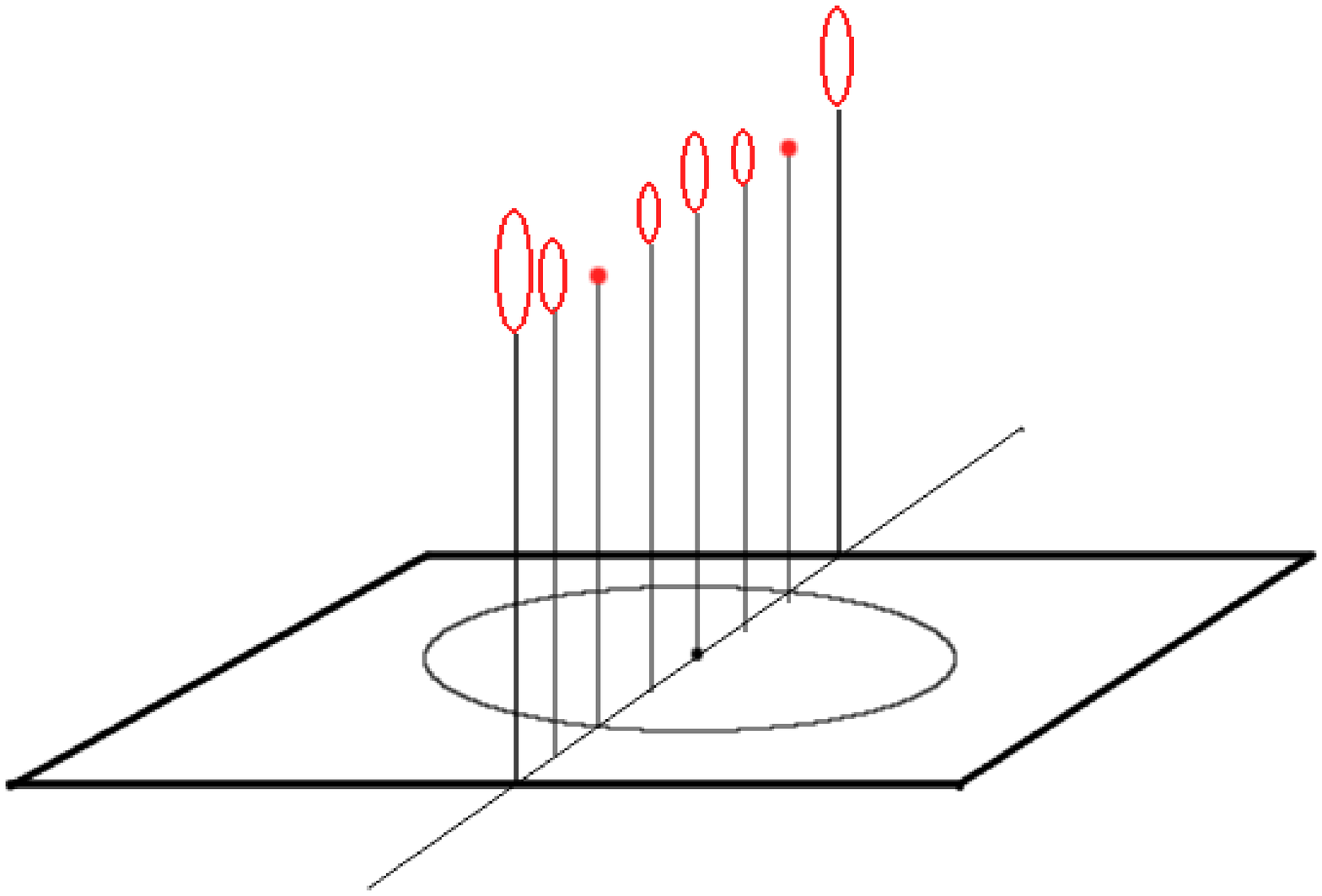}}}
\put(8,0) {\tfbox{\includegraphics[width=0.45\textwidth,angle=0]{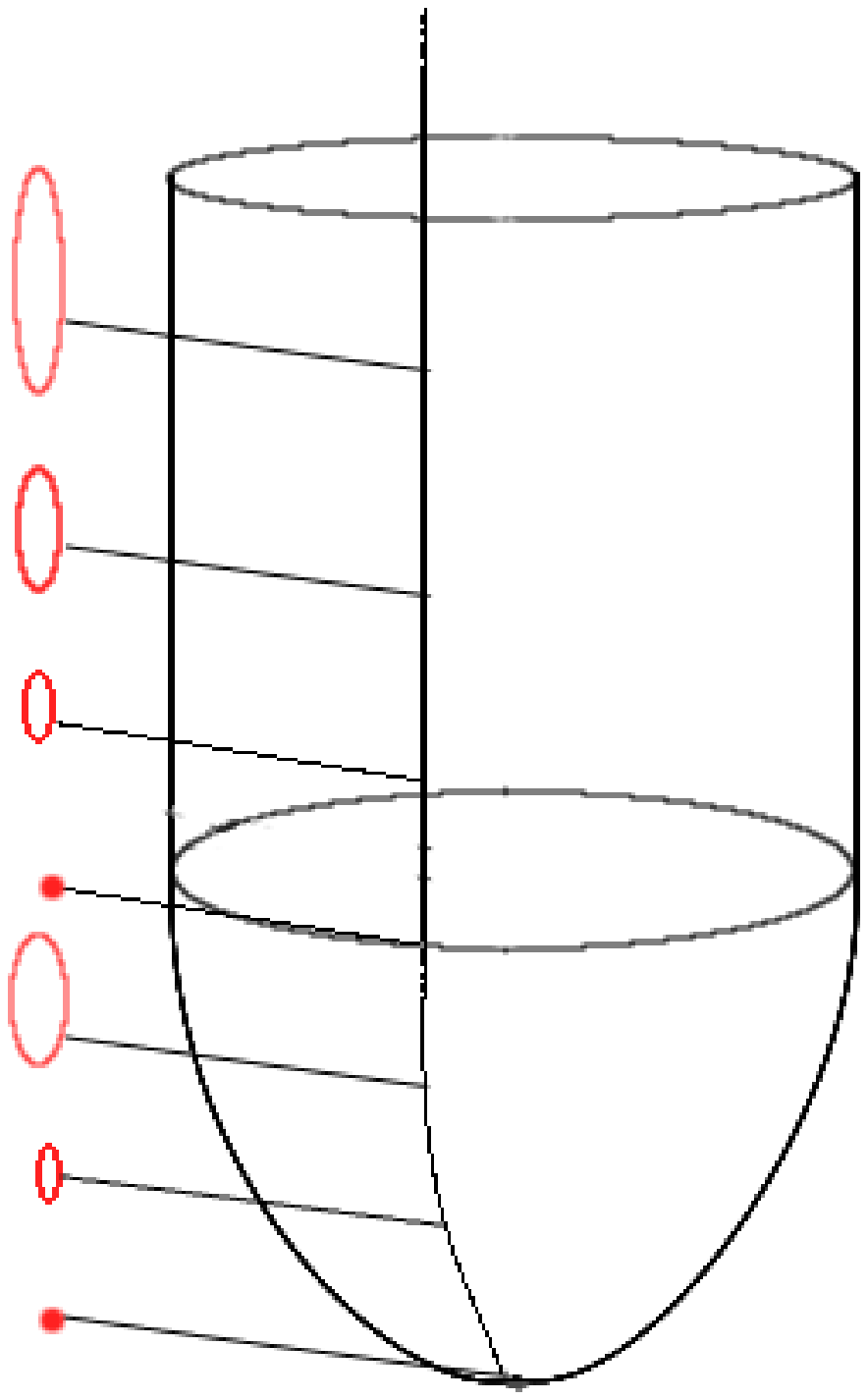}}}
\put(4.5,4.5) {$S^4$}
\put(9.3,4.5) {$S^4$}
\put(12,4.68) {$S^5$}
\end{picture}
\caption{\label{pinchpict} Left: The black circle represents a KK monopole supertube   with  a circular profile of radius $a$ in 5 dimensions. At every point of the
curve, the internal circle $S^4$ (drawn in red) pinches off to zero size. Right: After placing another KK monopole
wrapped on $S^4$
in the origin, the asymptotic geometry becomes ${\bf R}^4 \times S^5$. As argued in the text, the $S^4$ circle pinches off along the curve as well as in the origin. }
\end{figure}}

These are precisely the solutions constructed by Bena and Kraus
\cite{Bena:2005ay}\footnote{To make contact with the conventions
in \cite{Bena:2005ay}, one has to make a further coordinate transformation
$\f \to -\f, \theta \to \p - \theta$.}. They represent Kaluza-Klein
monopole supertubes which have been embedded into a Taub-NUT space
which has the asymptotic spatial geometry  ${\bf R}^3 \times S^5$.
By varying the radius $R_5$ of the circle $S^5$ we can interpolate
between solutions in 4 and in 5 noncompact dimensions; this
procedure goes under the name of the `4D-5D connection'
\cite{Gaiotto:2005gf,Gaiotto:2005xt}. The 5D solutions one obtains
in this way are highly symmetric fuzzball solutions where the
curve that defines the supertube is circular.

\subsection{4D-5D connection and 5D  fuzzball geometries}
Let us illustrate this in more detail. We take the
$R_5 \to \infty$ limit keeping the following quantities fixed:
\be
 2 r R_5 \equiv \tilde r^2 \,, \qquad 2 a R_5 \equiv \tilde a^2/n^2\,.\label{4D5Dlimit}
\ee
After taking this limit, the $p^1$ charge $N_K$ of our configuration becomes a deficit angle
and one obtains a configuration embedded in an   orbifold space ${\bf R}^4/{\bf Z}_{N_K}$.
We will therefore specialize to the case $N_K=1$  from now on, so that we obtain solutions
in asymptotically flat space.
We  define charges $\tilde Q_1 , \tilde Q_5 $ which remain finite in the limit (\ref{4D5Dlimit}) and
are the correctly normalized D1 and D5-brane charges in 5 noncompact dimensions:
\bea
\tilde Q_1 &=& 2 R_5 Q_1 = { g (2 \p)^4 \a'^3 N_1 \over V_{T_2 \times T_3}}\,,\nonu
\tilde Q_5 &=& 2 R_5 Q_1 = g \a' N_5\,.
\eea
The constraint (\ref{distconstr1})  on the distance between the centers  then reduces to
\be R_4 = {\sqrt{ \tilde Q_1 \tilde Q_5}\over \tilde a}\,.\label{posconstr5D}\ee
 The solution (\ref{metricclassI}, \ref{3formclassI})
can, in this limit, be written as a  fuzzball solution with a circular profile function \cite{Maldacena:2000dr,Lunin:2001jy,Lunin:2002qf,Lunin:2002iz}:
\bea
ds^2 &=& { 1 \over \sqrt { H^2 H^3} } \left[ - (dt + k)^2  + ( dx^4 - s -  k)^2\right]
+ \sqrt { H^2 H^3} d{\bf x}^2 + \sqrt{ H^2 \over H^3} ds^2_{T_2 \times T_3}\,,\nonu
e^{2 \F} &=& {H^2 \over H^3}\,,\nonu
F^{(3)} &=& d \left[ { 1\over  H^2 } (dt + k) \wedge ( dx^4 -s-k )\right] - \star_4 d ( H^3)\,,
\eea
where the harmonic functions are given by
\bea
 H^2 &=& 1 + {\tilde Q_5 \over L} \int_0^L {d v \over |{\bf x} - {\bf F}|^2}\,, \nonu
 H^3 &=& 1 + {\tilde Q_5 \over L} \int_0^L {|\dot {\bf F}|^2 d v \over |{\bf x} - {\bf F}|^2}\,,
\eea
and the one-foms $k,s$ take the form
\bea
s &=& {\tilde Q_5 \over L} \int_0^L {d v F^a \over |{\bf x} - {\bf F}|^2 } d x^a\,,\nonu
d(s + k) &=& - \star_4 d s\,. \label{5Dsol}
\eea
Here, ${\bf x}$ represents Cartesian coordinates on ${\bf R}^4$ which, in terms of the
coordinates  $ \tilde r, \theta, \f, \psi$ introduced earlier, are given by
\be
\begin{array}{lcl}
x^1 = \tilde r \cos {\theta \over 2} \cos \left( \psi + {\f\over 2}\right)\,,&\qquad & x^3 = \tilde r \sin {\theta \over 2} \cos \left( \psi - {\f\over 2}\right) \,,\\
x^2 = \tilde r \cos {\theta \over 2} \sin \left( \psi + {\f\over 2}\right) \,,& \qquad &x^3 = \tilde r \sin {\theta \over 2} \sin \left( \psi - {\f\over 2}\right)\,.
\end{array}
\label{cartcoords}
\ee
The profile function ${\bf F}(v)$ describes a circular profile in the $x^1-x^2$ plane:
\be
\begin{array}{lcl}
F^1 = {\tilde a \over n}\, \cos {2 \p n \over L} v ,& \qquad &F^3  = 0\,,\\
F^2 = {\tilde a \over n}\, \sin {2 \p n \over L} v, & \qquad &F^4
=0 \,.
\end{array}\label{circprofile}
\ee
 where
$ L \equiv {2 \p \tilde Q_5 \over R_4} $. The averaged length of
the tangent vector to the profile should be proportional to the
D1-brane charge: \be Q_1 = {Q_5 \over L} \int_0^L |\dot {\bf F}|^2
dv\,. \ee As a consistency check, one can easily see that this is
the case using the constraint (\ref{posconstr5D}).

Let us also discuss how the 5D angular momenta are related to quantum numbers in 4D.
Solutions in five noncompact dimensions can have 2 independent angular momenta $J_{12}$ in the
$x^1-x^2$-plane and $J_{34}$ in the $x^3-x^4$-plane.
These are related to the $R$-symmetry generators $J^3$ and $\bar J^3$ in the dual CFT as
$J_{12} = - (J^3 + \bar J^3),\  J_{12} = - (J^3 - \bar J^3)$. From the parametrization (\ref{cartcoords})
we see that $J^3$ comes from a linear momentum in four dimensions while $\bar J^3$ is proportional to
the four-dimensional angular momentum $J_z$.
This leads to the dictionary between the charges that was anticipated in \ref{chargedict}.
More specifically,
the solutions
above have
$ J_{12} = {N_1 N_5 \over n}, J_{34} = 0 $, so that
\be J^3 = \bar J^3 = -  {N_1 N_5 \over 2 n} .\ee

\subsection{Spectral flow and fuzzball solutions with momentum}
In paragraph \ref{spectflow}, we considered solutions that were obtained by a spectral flow transformation
labeled by a parameter $m$
that had the effect of adding D0-charge (\ref{totchargeclass2}). In the dual frame B, these will carry nonzero momentum charge P on the $S^4$ circle.
The harmonic functions and constraint on the distance were given in (\ref{harmfctnsclass2}, \ref{distconstrclass2}).
When we take the special case $Q_1 = Q_5$, substituting in (\ref{6Dform}) gives a solution with constant dilaton
which can be embedded in minimal 6-dimensional supergravity \cite{Gutowski:2003rg}. This solution  precisely matches the solutions
constructed in \cite{Saxena:2005uk} representing fuzzball geometries with momentum placed in a Taub-NUT space.

We can again take the  5D limit $R_5 \to \infty$ as discussed above.
Taking again $N_K=1$ to
get solutions in flat space,
 one obtains the five-dimensional fuzzball solutions with momentum that were
constructed in \cite{Lunin:2004uu,Giusto:2004ip,Giusto:2004id,Giusto:2004kj}.
These solutions  were originally obtained by applying a spectral flow transformation
to the five-dimensional solutions without momentum (\ref{5Dsol}).
They carry the following 5D charges
\be \begin{array}{lcl}
J^3 = - {N_1 N_5 \over 2} \left( 2 m + {1\over n} \right)\,, & \qquad & \bar J^3 = -{N_1 N_5 \over 2 n}\,,\\
P = N_1 N_5 m \left(m+ {1 \over n}\right),& &
\end{array}
\ee
where $P$ denotes the momentum on the $S^4$ circle. The flux quantization discussed in paragraph \ref{spectflow} imposes
that the parameter $m$ should be an integer.

\section{Microscopic interpretation}\label{micro}

We will now discuss the microscopic interpretation of the solutions we considered both from the 4D and 5D point of view.
Let us start with the configurations (\ref{chargesclass1}) without D0-charge in  frame A.
We showed that these arise, through the 4D-5D connection, from 5D fuzzball solutions with  circular profile which  carry macroscopic
angular momentum   $J_{12} =  N_1 N_5 / n $ and are placed in a Taub-NUT geometry.
 A first question is whether we should regard these solutions as
zero-entropy constituents of a spinning  black hole or of a  black ring  in five dimensions.
In the present context, the latter is the only possibility,
since a
black hole of the desired charge  (if it exists as a BPS solution in type II on a torus) cannot be placed in Taub-NUT space in a supersymmetric manner and therefore
the 4D-5D connection cannot be applied to it. Indeed, if it could, the resulting 4D configuration would be a small black
hole with charges $(0,N_K, N_1, N_5,  N_1 N_5/n,0,0,0)$. This is however a polar charge for which there cannot
exist a single center black hole solution,  even including higher derivative corrections. Hence
 we should see our 4D solutions as coming from small black ring microstates in five dimensions. This
interpretation also corresponds
to the one argued   in \cite{Lunin:2002qf,Iizuka:2005uv,Dabholkar:2005qs,Balasubramanian:2005qu}.
We want to point out that the above argument does not rule out the existence of a 5D supersymmetric spinless ($J_{12} = J_{34} =0$) small black hole placed
at the center of Taub-NUT space. Indeed, the resulting 4D configuration would have pure D4-charge $(0,N_K, N_1, N_5, 0,0,0,0)$, which is not
a polar charge ($\hat q_0 = 0$), and therefore could give rise to a single-centered small black hole when higher
derivative corrections are taken into account.

Let us review which  states in the dual CFT correspond to the configurations (\ref{chargesclass1})  from the 5D point of view.
The D1-D5 CFT is a deformation of a symmetric product  CFT with target space $({T_2 \times T_3})^{N_1 N_5} / S_{N_1 N_5}$ (see \cite{David:2002wn} for a review).
For our purposes, we can consider the theory at the orbifold point. The states we are considering are closely
related to chiral primary operators denoted by $\s_n^{--}$ with quantum numbers $L_0 = J^3 = \bar L_0 = \bar J^3 = {n-1\over 2}$.
We can construct operators  $U(\a)$ which generate a left-moving spectral flow with an integer parameter $\a$:
\bea
U(\a) L_0 U(\a)^{-1} &=& L_0 - \a J^3 +\a^2 {c \over 24} \nonu
U(\a) J^3 U(\a)^{-1} &=& J^3 - \a {c \over 12}
\eea
where the  central charge is  $c = 6 N_1 N_5$. Similar generators of right-moving spectral  flow with parameter
$\tilde \a$ will be denoted by $\tilde U (\tilde \a)$.
The CFT  states corresponding to (\ref{chargesclass1}) are ground states in the R sector given by
\be
U(1)\tilde U(1)( \s_n^{--} )^{N_1 N_5 \over n}|0\rangle \label{CFTstatesclass1}.
\ee
They carry the quantum numbers
\be \begin{array}{lcl}
L_0 = {N_1 N_5\over 4}\,, & \qquad & \bar L_0 = {N_1 N_5\over 4}\,, \\
J^3 = - {N_1 N_5 \over 2 n} \,, & \qquad & \bar J^3 = -{N_1 N_5 \over 2 n}\,,\\
P = L_0 - \bar L_0 = 0\,. & &
\end{array}
\ee

The above states belong to a `microcanonical' ensemble of R ground states at fixed
D1-charge $N_1$, D5-charge $N_5$, and angular momenta\footnote{A different ensemble, where the angular momenta are not fixed, was advocated in the light
of the OSV conjecture in \cite{Dabholkar:2005qs}} $J_{12} = N_1 N_5/n,\ J_{34} = 0$.
When $n\gg 1$, $J_{12}$ is sufficiently far from the
maximal value $N_1 N_5$, and there is  an exponential degeneracy of
states carrying these quantum numbers, leading to a microscopic
entropy \cite{Balasubramanian:2005qu} \be S_{\rm micro} = 2\sqrt{2}\p \sqrt{N_1 N_5 - J } =
2\sqrt{2}\p \sqrt{N_1 N_5(1 - {1 \over n})}\,.\label{Smicro} \ee
It is expected on the basis of general arguments \cite{Sen:1995in} that,
after including higher derivative corrections to the effective
action,  there exists a black ring solution with a matching
macroscopic entropy. It is an open problem to explicitly
compute such corrections in toroidal compactifications, unlike
the case where  the four-torus $T_2 \times T_3$
is  replaced with  $K_3$
\cite{Iizuka:2005uv,Dabholkar:2005qs,Dabholkar:2006za}.

When a small black ring is placed in Taub-NUT space with one unit of NUT charge and the radius of the Taub-NUT circle is decreased,
one obtains a 4D configuration consisting of two centers. One center, coming from the wrapped ring itself,
 becomes a small black hole in 4D, while the other center,  coming from the Taub-NUT charge,
is a KK monopole carrying zero entropy \cite{Iizuka:2005uv,Dabholkar:2005qs}.
In our duality frame A, the first center is a small $D4-D2$ black hole with charge $(0,0, N_1, N_5, N_1 N_5/n,0,0,0)$ and entropy given by (\ref{Smicro}) and
the second center is a pure D4-brane with charge $(0,1,0,0,0,0,0,0)$. Because these
charges are not parallel, the combined system carries macroscopic angular momentum $J_z = -  N_1 N_5/n$.
  Therefore we can see our 4D polar D6-anti D6 configurations (\ref{chargesclass1}) as zero-entropy constituents of
this two-centered configuration.

A similar discussion can be made for the solutions (\ref{chargesclass2})  carrying D0-charge in frame A. Their
CFT counterparts are related to (\ref{CFTstatesclass1}) by an additional left-moving spectral flow with
parameter $2m$:
\be
U(2m + 1)\tilde U(1)( \s_n^{--} )^{N_1 N_5 \over n}|0\rangle \label{CFTstatesclass2}.
\ee
They carry the quantum numbers that were anticipated in (\ref{CFTqn}):
\be \begin{array}{lcl}
L_0 = N_1 N_5 \left(m^2 + {m\over n} + 1/4\right)\,, & \qquad & \bar L_0 = {N_1 N_5\over 4}\,, \\
J^3 = - {N_1 N_5 \over 2} \left( 2 m + {1\over n} \right)\,, & \qquad & \bar J^3 = -{N_1 N_5 \over 2 n}\,,\\
P = L_0 - \bar L_0 = N_1 N_5 m \left(m+ {1 \over n}\right)\,. & &
\end{array}
\ee
In the CFT, the parameters $n$ and $m$ should be quantized such that $n$ is a divisor of $N_1 N_5$ and $m$
is an integer. This matches with the  conditions we found from charge quantization in the
corresponding D-brane configurations.
These states are part of an ensemble of CFT states with fixed D1-D5 charges, angular momenta $J^3, \bar J^3$ and momentum $P$.
This ensemble is obtained by the ensemble of zero momentum ground states discussed above by acting with
the spectral flow operator $U(2m)$. The degeneracy is then again given by (\ref{Smicro}).

\section{Discussion}\label{concl}

In this paper we have identified four-dimensional multicenter D-brane configurations that correspond
to a class of fuzzball solutions in five noncompact dimensions under the 4D-5D connection. In a
type IIA duality frame where all the charges come from D6-D4-D2-D0 branes, the relevant 4D configurations are
two-centered D6-anti D6 solutions with fluxes  corresponding to polar states .

The fuzzball solutions considered here were highly symmetric, where the
 profile function that defines the solution is taken to be a circular
curve in the  $x^1-x^2$ plane in the coordinates (\ref{cartcoords}). Let us
first comment on the fate of more general fuzzball solutions under the
4D-5D connection. A fuzzball solution arising from a generic curve
will typically not have enough symmetry to be written as a torus fibration
over a four-dimensional base as in (\ref{10Dlift}) and can hence not be
given a four-dimensional interpretation. However, according to the proposed
dictionary between microstates and fuzzball solutions  in \cite{Skenderis:2006ah,Kanitscheider:2006zf},
the subclass of fuzzball solutions that semiclassically represent eigenstates of the R-symmetry group
should possess $U(1)\times U(1)$ symmetry and be represented by (possibly disconnected)
circular curves in the $x^1-x^2$ and $x^3-x^4$ planes in the coordinates (\ref{10Dlift}). Such solutions
have isometries along the directions $\pa / \pa \f$ and $\pa / \pa  \psi $ as well as
along the Taub-Nut direction $\pa / \pa x^4$, and should therefore be the lift of axially symmetric solutions
in four dimensions. When the quantum numbers are chosen appropriately, these would describe
other constituents of the 4-dimensional 2-centered system with entropy (\ref{Smicro}). It would be interesting to explore this ensemble of four-dimensional
configurations.

We would also like to comment on the relation between  the present work and black hole deconstruction \cite{Denef:2007yt}. In four dimensions, say in our frame A, there exist multicentered `scaling' solutions with  centers  so close  that their throats have `melted' together and which are asymptotically  indistinguishable from single centered solutions. Such solutions can carry the same charges as a large single-centered D4-D0 black hole, and can be seen as a  deconstruction of such a black hole into zero-entropy constitutents. The scaling solutions
consist of a `core' D6 anti-D6 system with flux, and a `halo'of D0-brane centers added to it (again, see \cite{Denef:2007vg} for more details on the formalism of `cores' and `halos').
The scaling limit consists of taking the total D0-charge to be parametrically larger than the magnetic charge
$p^1 p^2 p^3$. The entropy of the black hole  in this limit can be understood by treating the D0-branes as probes
and counting the supersymmetric ground states of the probe quantum mechanics \cite{Gaiotto:2004ij}.
The `core' D6 anti-D6 system in these configurations is precisely of the kind  that we studied
in this paper and mapped to 5D fuzzball solutions. Indeed, for the special values $n=1,\  m = -1/2 $ of our parameters we obtain the following charges at the centers
\bea \G_1&=& \left(-1,\frac{N_K}{2},\frac{N_1}{2},\frac{N_5}{2},-\frac{N_1N_5}{4},-\frac{N_KN_5}{4},-\frac{N_KN_1}{4},
\frac{N_KN_1N_5}{8}\right),\nonumber\\
\G_2&=&\left(1,\frac{N_K}{2},\frac{N_1}{2},\frac{N_5}{2},\frac{N_1N_5}{4},\frac{N_KN_5}{4},\frac{N_KN_1}{4},
\frac{N_KN_1N_5}{8}\right)\,.
\eea
These are precisely  the charges that appear in the core of the scaling solutions in \cite{Denef:2007yt}. It seems
natural to expect that, for the other values of our parameters $m$ and $n$, our configurations can serve as
the core system for the deconstruction of a black hole with added D2-charge.

The relation to deconstruction could have interesting implications in five dimensions as well. If we take
a scaling solution in four dimensions, dualize it to  frame B and take the 4D-5D limit, we should end
up with a  configuration carrying the charges of a large D1-D5-P Strominger-Vafa \cite{Strominger:1996sh} black hole.
The scaling limit implies that we will have $P \gg N_1 N_5$, which is equivalent to the Cardy
limit $\L_0 \ll c$  where the CFT microstate counting is performed. Therefore such configurations
would be candidates for describing typical microstates of the D1-D5-P black hole, and
it would be interesting
to study such solutions in more detail.
It is not clear whether such configurations could rightly be called `fuzzball' geometries for the D1-D5-P black hole,
as they will not be smooth near the centers where the harmonic functions describing the momentum diverge.
As argued in \cite{Raeymaekers:2007ga}, treating the momentum as  coming from giant graviton probes,
the number of ground states would be of the right order to explain the entropy.

\acknowledgments{We would like to thank  Bram Gaasbeek for initial collaboration and  Joke Adam,
Andres Collinucci, Frederik Denef, Laura Tamassia and Dieter Van den Bleeken for useful discussions.
This work  is supported in part by the European Community's Human Potential
Programme under contract MRTN-CT-2004-005104 `Constituents,
fundamental forces and symmetries of the universe', in part by the FWO-Vlaanderen, project G.0235.05 and in part by the Federal Office for Scientific,
Technical and Cultural Affairs through the ‘Interuniversity Attraction
Poles Programme – Belgian Science Policy’ P6/11-P. B.V. is aspirant FWO-Vlaanderen.}

\begin{appendix}

\section{Reduction formulas in frame B}\label{reductionBframe}

We now discuss the dimensional reduction of type II on $T^6$ in the duality frame B
to the bosonic STU model action (\ref{STUaction}). The 10-dimensional interpretation of the
$U(1)$ charges is given in table \ref{frameBcharges}.
It will be convenient to first reduce to an intermediate duality frame, which we will call frame $\tilde{B}$, where
the $U(1)$ fields are labeled as $\ca^0,\ca^1,\cb_2,\ca^3$ and the charges are labeled as
$(p^0,p^1,\tilde p^2,p^3, q_1,\tilde q_2,q_1,q_0)$. The 10D interpretation of the charges in frame $\tilde{B}$ is
given in table \ref{frameBtilde}.
\TABLE{}{
\begin{table} \begin{center}
\begin{tabular}{cl|cl}
$q_0$& $P(S^4) $       & $p^0$& $KKmon(S^4)$\\
$q_1$& $P( S^5)$ & $p^1$& $KKmon( S^5)$\\
$\tilde q_2$& $D1(S^4)$       & $\tilde p^2$& $D5(S^4)$\\
$q_3$& $D1( S^5)$& $p^3$& $D5( S^5)$\\
\end{tabular}
\end{center}
\caption{\label{frameBtilde}The interpretation of the charge in an intermediate frame $\tilde B$.}
\end{table}}
The frame $\tilde{B}$ differs from the frame B of table \ref{frameBcharges} by an electromagnetic duality transformation on
the $U(1)$ field $\cb^2$.

It suffices to restrict attention to a truncated IIB action containing only the metric, dilaton and RR 3-form:
\be S = {1 \over (2 \p)^7 \a'^4} \int d^{10}x  \sqrt{-G^{(10)}} \left[ e^{-2 \F^{(10)}} \left( R^{(10)}
+ 4 \pa_M \F^{(10)} \pa^M \F^{(10)}\right) - {1 \over 12} F_{MNP}^{(10)}F^{(10)\; MNP} \right]\,.\label{10Daction}\ee
We perform a trivial dimensional reduction over the four-torus $T_2 \times T_3$, while allowing the
torus $T_1$ to be nontrivially fibered over the four-dimensional base. We start by flipping the sign
of $\F^{(10)}$ and making a Weyl transformation (as one does in S-duality) such that all terms
in   (\ref{10Daction}) have an $ e^{-2 \F^{(10)}}$
factor in front. We can then perform the dimensional reduction of this sector as discussed in  \cite{Maharana:1992my}. We will here follow closely the conventions of \cite{Sen:1994fa}.
 We take the following
reduction ansatz
\bea \label{10dtable}
\F^{(10)} &=& - \F - { 1 \over 4}\ln \det \hat{G}_{mn}- { 1 \over 4}\ln \det \hat{G}_{ij}\,,\nonu
G^{(10)}_{\m\n} &=& ( \det \hat{G})^{-1/4} \left( e^\F G_{\m\n} + 2\norm^2   e^{- \F}  \ca^{m-4}_\m
\ca^{n-4}_\n
\hat{G}_{mn} \right)\,,\nonu
G^{(10)}_{\m n } &=& \sqrt{2  }\norm  ( \det \hat{G})^{-1/4}  e^{-\F} \hat{G}_{np}  \ca^{p-4}_\m \,,\nonu
G^{(10)}_{m n } &=& ( \det \hat{G})^{-1/4}  e^{-\F} \hat{G}_{mn}\,,\nonu
G^{(10)}_{ij} &=& ( \det \hat{G})^{-1/4}  e^{-\F}\hat{G}_{ij}\,,\nonu
C^{(10)}_{\m\n} &=& C_{\m\n}  + 2   \norm^2 \hat{C}_{45} ( \ca^{0}_\m  \ca^{1}_\n -
 \ca^{1}_\m  \ca^{0}_\n) + \norm^2 ( \ca^{0}_\m  \cb_{2\n} -
 \cb_{2\m} \ca^{0}_\n)+  \norm^2 ( \ca^{1}_\m  \ca^{3}_\n -
 \ca^{3}_\m  \ca^{1}_\n)\,,\nonu
C^{(10)}_{\m 4} &=& \sqrt{2 } \norm( \cb_{2\m} +  \hat{C}_{45}  \ca^{1}_\m)\,,\nonu
C^{(10)}_{\m 5} &=&   \sqrt{2  } \norm^2 (\ca^{3}_\m -  \hat{C}_{45}  \ca^{0}_\m)\,,\nonu
C^{(10)}_{m n} &=& \hat{C}_{mn}\,.
\eea
Here, $M,N = 0, \ldots, 9;\ m,n = 4,5,\ i,j = 6,\ldots 9$  and we have taken $x^4, x^5$ to parametrize $S_4,  S_5$ respectively.

The matrix $\hat{G}_{ij}$ is a constant metric on $T_2 \times T_3$ and the matrices $\hat{G}_{mn},\hat{C}_{mn}$
can be conveniently parametrized as
\bea
\hat{G}_{mn} &=&  b_3  \left( \matrix{{a_1^2 + b_1^2\over b_1} & - {a_1 \over b_1}\cr  -{a_1 \over b_1}& {1\over b_1} } \right)\,,\nonu
\hat{C}_{mn} &=&  \left( \matrix{0 & a_3 \cr - a_3 & 0 } \right)\,,\nonu
e^{-2\F} &=&  b_2\,.
\eea
The two-form $C_{\m\n}$  can be dualized in four dimensions to give another scalar  $\tilde a_1$:
\be d a_2 = b_2^2 \star F\,, \ee where the Hodge $\star$ is to be
taken with respect to the 4D metric $G_{\m\n}$ and the three-form
field strength $F$ is defined as \be F = d C +  {\norm^2 \over 2}\left( \ca^0 \wedge
\cg^2 +   \cb^2 \wedge  \cf^0 + \ca^1 \wedge  \cf^3 +  \ca^3
\wedge \cf^1\right)  \,.\ee
{}From the above expressions it is clear that $z^1 = a_1 + i b_1$ is the complex structure modulus of $T_1$,
$z^2 = a_2 + i b_2$ is the 4D axion-dilaton and $z^3 = a_3 + i b_3$ is the complexified K\"ahler modulus of $T_1$.

In these variables, one finds
after performing the dimensional  reduction  the 4D action
\bea S &=& {1\over 16 \p
G_4 } \int d^4 x \sqrt{-G}\Big{[} R- 2 \sum_{A=1}^3 {\pa_\m \tilde z^A
\pa^\m \bar {\tilde z}^A\over (\tilde z^A - \bar {\tilde z}^A)^2} \nonu
& &+ {\norm^2  \over 2 } {\rm Im}
\tilde \cn_{IJ} \cf^I_{\m\n}\cf^{J\;\m\n}  + {\norm^2 \over 4 } {\rm Re}
\tilde \cn_{IJ}\e^{\m\n\r\s}
\cf^I_{\m\n}\cf^J_{\;\r\s}\Big{]},\,.\eea
with the matrix $\tilde \cn$ given by
{\small \bea
{\rm Re} \tilde \cn &=&   \left(
\begin{array}{llll}
0&0&-a_2&0\\ 0&0&0&-a_2\\-a_2&0&0&0\\0&-a_2&0&0
\end{array}
\right)\,,\nonu
{\rm Im} \tilde \cn &=&   \left(
\begin{array}{llll}
 -\frac{\text{b_2} \left(\text{a_1}^2+\text{b_1}^2\right)
   \left(\text{a_3}^2+\text{b_3}^2\right)}{\text{b_1} \text{b_3}} & \frac{ \text{a_1} \text{b_2}
   \left(\text{a_3}^2+\text{b_3}^2\right)}{\text{b_1} \text{b_3}} & -\frac{ \text{a_1}
   \text{a_3} \text{b_2}}{\text{b_1} \text{b_3}} & \frac{ \text{a_3} \text{b_2}
   \left(\text{a_1}^2+\text{b_1}^2\right) }{\text{b_1} \text{b_3}} \\
 \frac{ \text{a_1} \text{b_2} \left(\text{a_3}^2+\text{b_3}^2\right)}{\text{b_1} \text{b_3}} & -\frac{
   \text{b_2} \left(\text{a_3}^2+\text{b_3}^2\right)}{\text{b_1} \text{b_3}} & \frac{ \text{a_3}
   \text{b_2}}{\text{b_1} \text{b_3}} & -\frac{ \text{a_1} \text{a_3} \text{b_2}}{\text{b_1}
   \text{b_3}} \\
 -\frac{ \text{a_1} \text{a_3} \text{b_2}}{\text{b_1} \text{b_3}} & \frac{ \text{a_3}
   \text{b_2}}{\text{b_1} \text{b_3}} & -\frac{ \text{b_2}}{\text{b_1} \text{b_3}} & \frac{ \text{a_1}
   \text{b_2}}{\text{b_1} \text{b_3}} \\
 \frac{ \text{a_3}  \text{b_2} \left(\text{a_1}^2+\text{b_1}^2\right)}{\text{b_1} \text{b_3}} &
   -\frac{ \text{a_1} \text{a_3} \text{b_2}}{\text{b_1} \text{b_3}} & \frac{ \text{a_1}
   \text{b_2}}{\text{b_1} \text{b_3}} & -\frac{  \text{b_2} \left(\text{a_1}^2+\text{b_1}^2\right)
  }{\text{b_1} \text{b_3}}
\end{array}\,.
\right)\nonumber \eea} The 4-dimensional Newton constant $G_4$ is
given by \be G_4 = { 8 \p^6 (\a ')^4 g^2 \over (2\p)^2 R_4 R_5
V_{T_2 \times T_3}}\,,\ee with $g$ the string coupling in 10
dimensions.

To go to the duality frame B of table \ref{frameBcharges}, where
the $U(1)$ fields are labeled as $\ca^0,\ca^1,\cb^2,\ca^3$ and the charges are labeled as
$(p^0,p^1, p^2,p^3, q_1, q_2,q_3,q_0)$, we have to perform an electromagnetic duality
on the field $\cb_2$. After this duality, the action takes the form (\ref{STUaction}) with the matrix $\cn$
related to $\tilde \cn$ given above by a symplectic transformation
\be \cn = ( C + D\tilde \cn) (A + B \tilde \cn)^{-1}\,,\label{sympltransfN}\ee
with
\be A = D = \left( \matrix{ 1&0&0&0\cr 0&1&0&0\cr 0&0&0&0\cr 0&0&0&1 }\right); B = -C =
\left( \matrix{ 0&0&0&0\cr 0&0&0&0\cr 0&0&-1&0\cr 0&0&0&0 }\right) \,.\ee
Explicitly, one finds
 {\small
\begin{eqnarray}
{\rm Re}{\cal N}\, &=&\, - \left(\matrix{ 2\,{a_1}\,{a_2}\,
   {a_3} & -\left( {a_2}\,{a_3}
      \right)  & -\left( {a_1}\,{a_3} \right)
      & -\left( {a_1}\,{a_2} \right)
      \cr -\left( {a_2}\,{a_3} \right)
      & 0 & {a_3} & {a_2} \cr -\left(
     {a_1}\,{a_3} \right)  & {a_3
   } & 0 & {a_1} \cr -\left( {a_1}\,
     {a_2} \right)  & {a_2} & {a_1
   } & 0 \cr  }\right)\,,\nonumber\\
{\rm Im}{\cal N}\, &=&\, - \left(\matrix{ b_1b_2b_3 + {b_1b_2a_3^2 \over b_3} +
{b_1b_3a_2^2 \over b_2}+ {b_2b_3a_1^2 \over b_1}
& -{ \frac{{a_1}\,{b_2}\,{b_3}}
     {{b_1}}} & -{\frac{{a_2}\,
       {b_1}\,{b_3}}{{b_2}}} &
    -{\frac{{a_3}\,{b_1}\,{b_2}}
     {{b_3}}} \cr -{\frac{{a_1}\,
       {b_2}\,{b_3}}{{b_1}}} &
    {\frac{{b_2}\,{b_3}}{{b_1}}} &
   0 & 0 \cr -{\frac{{a_2}\,{b_1}\,
       {b_3}}{{b_2}}} & 0 & {\frac{
      {b_1}\,{b_3}}{{b_2}}} & 0 \cr
   -{\frac{{a_3}\,{b_1}\,{b_2}}
     {{b_3}}} & 0 & 0 & {\frac{{b_1}\,
      {b_2}}{{b_3}}} \cr  }\right)\,.
\label{Ngen}\nonumber
\end{eqnarray}
} This is indeed the standard form of the matrix $\cn$ in the
STU-model derived from the prepotential through (\ref{NfromF}).
The $U(1)$ field $\cb_2$ is related to the $\ca^I$ through \be d
\cb_2 = {\rm Im}  \cn_{2J} \star \cf^J + {\rm Re}  \cn_{2J}
\cf^J\,.\ee

Summarized, we have found the following reduction formulas \bea
e^{2 \F^{(10)}} &=&  {b^2 \over  b^3}\,,\nonu ds^2_{10} &=& {1
\over \sqrt{ b^2  b^3} }ds^2_4 +  \sqrt{ b^2   b^3} \cm_{mn} (dx^m
+\sqrt{2 }\norm  \ca^{m-4})(dx^n + \sqrt{2 }\norm  \ca^{n-4}) + \sqrt{ b^2
\over  b^3}ds^2_{T_2 \times T_3}\,,\nonu \cm_{mn}  &=& {1 \over
b^1}\left( \matrix{ ( a^1)^2 +( b^1)^2 & - a^1 \cr - a^1 & 1}
\right)\,,\nonu C^{(10)} &=&  \half C_{\m\n} dx^\m dx^\n +  a^3
(dx^4 -\sqrt{2 } \norm \ca^{0})\wedge (dx^5 - \sqrt{2 }\norm \ca^{1}) \nonu &&
- \sqrt{2} \norm dx^4 \wedge  \cb^2  - \sqrt{2 }\norm dx^5 \wedge  \ca^3
+\norm^2 \left(\ca^0\wedge   \cb^2 + \ca^1 \wedge
 \ca^3\right) \,,\nonu
 d  a_2 &=& (b_2)^2 \star F \,,\nonu
F &=& d C + { \norm^2 \over 2 } (  \ca^0 \wedge  \cg^2 +   \cb^2 \wedge  \cf^0 +
\ca^1 \wedge  \cf^3 +  \ca^3 \wedge  \cf^1)\,.
\eea

\end{appendix}

\bibliography{ref}

\providecommand{\href}[2]{#2}\begingroup\raggedright\begin{thebibliography}{10}

\bibitem{Denef:2007vg}
F.~Denef and G.~W. Moore, ``{Split states, entropy enigmas, holes and halos},''
\href{http://arxiv.org/abs/hep-th/0702146}{{\tt arXiv:hep-th/0702146}}.

\bibitem{Denef:2000nb}
F.~Denef, ``{Supergravity flows and D-brane stability},'' {\em JHEP} {\bf 08}
  (2000)  050,
\href{http://arxiv.org/abs/hep-th/0005049}{{\tt arXiv:hep-th/0005049}}.

\bibitem{Denef:2001xn}
F.~Denef, B.~R. Greene, and M.~Raugas, ``{Split attractor flows and the
  spectrum of BPS D-branes on the quintic},'' {\em JHEP} {\bf 05} (2001)  012,
\href{http://arxiv.org/abs/hep-th/0101135}{{\tt arXiv:hep-th/0101135}}.

\bibitem{Bates:2003vx}
B.~Bates and F.~Denef, ``{Exact solutions for supersymmetric stationary black
  hole composites},''
\href{http://arxiv.org/abs/hep-th/0304094}{{\tt arXiv:hep-th/0304094}}.

\bibitem{deBoer:2008fk}
J.~de~Boer, F.~Denef, S.~El-Showk, I.~Messamah, and D.~Van~den Bleeken,
  ``{Black hole bound states in $AdS_3 x S^2$},''
\href{http://arxiv.org/abs/0802.2257}{{\tt arXiv:0802.2257 [hep-th]}}.

\bibitem{Ooguri:2004zv}
H.~Ooguri, A.~Strominger, and C.~Vafa, ``{Black hole attractors and the
  topological string},''
  \href{http://dx.doi.org/10.1103/PhysRevD.70.106007}{{\em Phys. Rev.} {\bf
  D70} (2004)  106007},
\href{http://arxiv.org/abs/hep-th/0405146}{{\tt arXiv:hep-th/0405146}}.

\bibitem{Denef:2007yt}
F.~Denef, D.~Gaiotto, A.~Strominger, D.~Van~den Bleeken, and X.~Yin, ``{Black
  hole deconstruction},''
\href{http://arxiv.org/abs/hep-th/0703252}{{\tt arXiv:hep-th/0703252}}.

\bibitem{Elvang:2004rt}
H.~Elvang, R.~Emparan, D.~Mateos, and H.~S. Reall, ``{A supersymmetric black
  ring},'' \href{http://dx.doi.org/10.1103/PhysRevLett.93.211302}{{\em Phys.
  Rev. Lett.} {\bf 93} (2004)  211302},
\href{http://arxiv.org/abs/hep-th/0407065}{{\tt arXiv:hep-th/0407065}}.

\bibitem{Bena:2004de}
I.~Bena and N.~P. Warner, ``{One ring to rule them all ... and in the darkness
  bind them?},'' {\em Adv. Theor. Math. Phys.} {\bf 9} (2005)  667--701,
\href{http://arxiv.org/abs/hep-th/0408106}{{\tt arXiv:hep-th/0408106}}.

\bibitem{Elvang:2004ds}
H.~Elvang, R.~Emparan, D.~Mateos, and H.~S. Reall, ``{Supersymmetric black
  rings and three-charge supertubes},''
  \href{http://dx.doi.org/10.1103/PhysRevD.71.024033}{{\em Phys. Rev.} {\bf
  D71} (2005)  024033},
\href{http://arxiv.org/abs/hep-th/0408120}{{\tt arXiv:hep-th/0408120}}.

\bibitem{Emparan:2006mm}
R.~Emparan and H.~S. Reall, ``{Black rings},'' {\em Class. Quant. Grav.} {\bf
  23} (2006)  R169,
\href{http://arxiv.org/abs/hep-th/0608012}{{\tt arXiv:hep-th/0608012}}.

\bibitem{Maldacena:2000dr}
J.~M. Maldacena and L.~Maoz, ``{De-singularization by rotation},'' {\em JHEP}
  {\bf 12} (2002)  055,
\href{http://arxiv.org/abs/hep-th/0012025}{{\tt arXiv:hep-th/0012025}}.

\bibitem{Lunin:2001jy}
O.~Lunin and S.~D. Mathur, ``{AdS/CFT duality and the black hole information
  paradox},'' \href{http://dx.doi.org/10.1016/S0550-3213(01)00620-4}{{\em Nucl.
  Phys.} {\bf B623} (2002)  342--394},
\href{http://arxiv.org/abs/hep-th/0109154}{{\tt arXiv:hep-th/0109154}}.

\bibitem{Lunin:2002qf}
O.~Lunin and S.~D. Mathur, ``{Statistical interpretation of Bekenstein entropy
  for systems with a stretched horizon},''
  \href{http://dx.doi.org/10.1103/PhysRevLett.88.211303}{{\em Phys. Rev. Lett.}
  {\bf 88} (2002)  211303},
\href{http://arxiv.org/abs/hep-th/0202072}{{\tt arXiv:hep-th/0202072}}.

\bibitem{Lunin:2002iz}
O.~Lunin, J.~M. Maldacena, and L.~Maoz, ``{Gravity solutions for the D1-D5
  system with angular momentum},''
\href{http://arxiv.org/abs/hep-th/0212210}{{\tt arXiv:hep-th/0212210}}.

\bibitem{Lunin:2004uu}
O.~Lunin, ``{Adding momentum to D1-D5 system},'' {\em JHEP} {\bf 04} (2004)
  054,
\href{http://arxiv.org/abs/hep-th/0404006}{{\tt arXiv:hep-th/0404006}}.

\bibitem{Giusto:2004ip}
S.~Giusto, S.~D. Mathur, and A.~Saxena, ``{3-charge geometries and their CFT
  duals},'' \href{http://dx.doi.org/10.1016/j.nuclphysb.2005.01.009}{{\em Nucl.
  Phys.} {\bf B710} (2005)  425--463},
\href{http://arxiv.org/abs/hep-th/0406103}{{\tt arXiv:hep-th/0406103}}.

\bibitem{Giusto:2004id}
S.~Giusto, S.~D. Mathur, and A.~Saxena, ``{Dual geometries for a set of
  3-charge microstates},''
  \href{http://dx.doi.org/10.1016/j.nuclphysb.2004.09.001}{{\em Nucl. Phys.}
  {\bf B701} (2004)  357--379},
\href{http://arxiv.org/abs/hep-th/0405017}{{\tt arXiv:hep-th/0405017}}.

\bibitem{Giusto:2004kj}
S.~Giusto and S.~D. Mathur, ``{Geometry of D1-D5-P bound states},''
  \href{http://dx.doi.org/10.1016/j.nuclphysb.2005.09.037}{{\em Nucl. Phys.}
  {\bf B729} (2005)  203--220},
\href{http://arxiv.org/abs/hep-th/0409067}{{\tt arXiv:hep-th/0409067}}.

\bibitem{Bena:2005ay}
I.~Bena and P.~Kraus, ``{Microstates of the D1-D5-KK system},''
  \href{http://dx.doi.org/10.1103/PhysRevD.72.025007}{{\em Phys. Rev.} {\bf
  D72} (2005)  025007},
\href{http://arxiv.org/abs/hep-th/0503053}{{\tt arXiv:hep-th/0503053}}.

\bibitem{Taylor:2005db}
M.~Taylor, ``{General 2 charge geometries},'' {\em JHEP} {\bf 03} (2006)  009,
\href{http://arxiv.org/abs/hep-th/0507223}{{\tt arXiv:hep-th/0507223}}.

\bibitem{Saxena:2005uk}
A.~Saxena, G.~Potvin, S.~Giusto, and A.~W. Peet, ``{Smooth geometries with four
  charges in four dimensions},'' {\em JHEP} {\bf 04} (2006)  010,
\href{http://arxiv.org/abs/hep-th/0509214}{{\tt arXiv:hep-th/0509214}}.

\bibitem{Giusto:2005ag}
S.~Giusto, S.~D. Mathur, and Y.~K. Srivastava, ``{Dynamics of supertubes},''
  \href{http://dx.doi.org/10.1016/j.nuclphysb.2006.07.029}{{\em Nucl. Phys.}
  {\bf B754} (2006)  233--281},
\href{http://arxiv.org/abs/hep-th/0510235}{{\tt arXiv:hep-th/0510235}}.

\bibitem{Giusto:2006zi}
S.~Giusto, S.~D. Mathur, and Y.~K. Srivastava, ``{A microstate for the 3-charge
  black ring},'' \href{http://dx.doi.org/10.1016/j.nuclphysb.2006.11.009}{{\em
  Nucl. Phys.} {\bf B763} (2007)  60--90},
\href{http://arxiv.org/abs/hep-th/0601193}{{\tt arXiv:hep-th/0601193}}.

\bibitem{Kanitscheider:2007wq}
I.~Kanitscheider, K.~Skenderis, and M.~Taylor, ``{Fuzzballs with internal
  excitations},'' {\em JHEP} {\bf 06} (2007)  056,
\href{http://arxiv.org/abs/0704.0690}{{\tt arXiv:0704.0690 [hep-th]}}.

\bibitem{Mathur:2005zp}
S.~D. Mathur, ``{The fuzzball proposal for black holes: An elementary
  review},'' \href{http://dx.doi.org/10.1002/prop.200410203}{{\em Fortsch.
  Phys.} {\bf 53} (2005)  793--827},
\href{http://arxiv.org/abs/hep-th/0502050}{{\tt arXiv:hep-th/0502050}}.

\bibitem{Skenderis:2008qn}
K.~Skenderis and M.~Taylor, ``{The fuzzball proposal for black holes},''
\href{http://arxiv.org/abs/0804.0552}{{\tt arXiv:0804.0552 [hep-th]}}.

\bibitem{Gaiotto:2005gf}
D.~Gaiotto, A.~Strominger, and X.~Yin, ``{New connections between 4D and 5D
  black holes},'' {\em JHEP} {\bf 02} (2006)  024,
\href{http://arxiv.org/abs/hep-th/0503217}{{\tt arXiv:hep-th/0503217}}.

\bibitem{Gaiotto:2005xt}
D.~Gaiotto, A.~Strominger, and X.~Yin, ``{5D black rings and 4D black holes},''
  {\em JHEP} {\bf 02} (2006)  023,
\href{http://arxiv.org/abs/hep-th/0504126}{{\tt arXiv:hep-th/0504126}}.

\bibitem{Elvang:2005sa}
H.~Elvang, R.~Emparan, D.~Mateos, and H.~S. Reall, ``{Supersymmetric 4D
  rotating black holes from 5D black rings},'' {\em JHEP} {\bf 08} (2005)  042,
\href{http://arxiv.org/abs/hep-th/0504125}{{\tt arXiv:hep-th/0504125}}.

\bibitem{Bena:2005ni}
I.~Bena, P.~Kraus, and N.~P. Warner, ``{Black rings in Taub-NUT},''
  \href{http://dx.doi.org/10.1103/PhysRevD.72.084019}{{\em Phys. Rev.} {\bf
  D72} (2005)  084019},
\href{http://arxiv.org/abs/hep-th/0504142}{{\tt arXiv:hep-th/0504142}}.

\bibitem{Behrndt:2005he}
K.~Behrndt, G.~Lopes~Cardoso, and S.~Mahapatra, ``{Exploring the relation
  between 4D and 5D BPS solutions},''
  \href{http://dx.doi.org/10.1016/j.nuclphysb.2005.10.026}{{\em Nucl. Phys.}
  {\bf B732} (2006)  200--223},
\href{http://arxiv.org/abs/hep-th/0506251}{{\tt arXiv:hep-th/0506251}}.

\bibitem{Ford:2007th}
J.~Ford, S.~Giusto, A.~Peet, and A.~Saxena, ``{Reduction without reduction:
  Adding KK-monopoles to five dimensional stationary axisymmetric solutions},''
  \href{http://dx.doi.org/10.1088/0264-9381/25/7/075014}{{\em Class. Quant.
  Grav.} {\bf 25} (2008)  075014},
\href{http://arxiv.org/abs/0708.3823}{{\tt arXiv:0708.3823 [hep-th]}}.

\bibitem{Bena:2005va}
I.~Bena and N.~P. Warner, ``{Bubbling supertubes and foaming black holes},''
  \href{http://dx.doi.org/10.1103/PhysRevD.74.066001}{{\em Phys. Rev.} {\bf
  D74} (2006)  066001},
\href{http://arxiv.org/abs/hep-th/0505166}{{\tt arXiv:hep-th/0505166}}.

\bibitem{Berglund:2005vb}
P.~Berglund, E.~G. Gimon, and T.~S. Levi, ``{Supergravity microstates for BPS
  black holes and black rings},'' {\em JHEP} {\bf 06} (2006)  007,
\href{http://arxiv.org/abs/hep-th/0505167}{{\tt arXiv:hep-th/0505167}}.

\bibitem{Bena:2006is}
I.~Bena, C.-W. Wang, and N.~P. Warner, ``{The foaming three-charge black
  hole},'' \href{http://dx.doi.org/10.1103/PhysRevD.75.124026}{{\em Phys. Rev.}
  {\bf D75} (2007)  124026},
\href{http://arxiv.org/abs/hep-th/0604110}{{\tt arXiv:hep-th/0604110}}.

\bibitem{Balasubramanian:2006gi}
V.~Balasubramanian, E.~G. Gimon, and T.~S. Levi, ``{Four Dimensional Black Hole
  Microstates: From D-branes to Spacetime Foam},''
  \href{http://dx.doi.org/10.1088/1126-6708/2008/01/056}{{\em JHEP} {\bf 01}
  (2008)  056},
\href{http://arxiv.org/abs/hep-th/0606118}{{\tt arXiv:hep-th/0606118}}.

\bibitem{Bena:2006kb}
I.~Bena, C.-W. Wang, and N.~P. Warner, ``{Mergers and typical black hole
  microstates},'' {\em JHEP} {\bf 11} (2006)  042,
\href{http://arxiv.org/abs/hep-th/0608217}{{\tt arXiv:hep-th/0608217}}.

\bibitem{Cheng:2006yq}
M.~C.~N. Cheng, ``{More bubbling solutions},'' {\em JHEP} {\bf 03} (2007)  070,
\href{http://arxiv.org/abs/hep-th/0611156}{{\tt arXiv:hep-th/0611156}}.

\bibitem{Bena:2007ju}
I.~Bena, N.~Bobev, and N.~P. Warner, ``{Bubbles on Manifolds with a U(1)
  Isometry},'' \href{http://dx.doi.org/10.1088/1126-6708/2007/08/004}{{\em
  JHEP} {\bf 08} (2007)  004},
\href{http://arxiv.org/abs/0705.3641}{{\tt arXiv:0705.3641 [hep-th]}}.

\bibitem{Gimon:2007mha}
E.~G. Gimon and T.~S. Levi, ``{Black Ring Deconstruction},''
  \href{http://dx.doi.org/10.1088/1126-6708/2008/04/098}{{\em JHEP} {\bf 04}
  (2008)  098},
\href{http://arxiv.org/abs/0706.3394}{{\tt arXiv:0706.3394 [hep-th]}}.

\bibitem{Bena:2007qc}
I.~Bena, C.-W. Wang, and N.~P. Warner, ``{Plumbing the Abyss: Black Ring
  Microstates},''
\href{http://arxiv.org/abs/0706.3786}{{\tt arXiv:0706.3786 [hep-th]}}.

\bibitem{Bena:2007kg}
I.~Bena and N.~P. Warner, ``{Black holes, black rings and their microstates},''
\href{http://arxiv.org/abs/hep-th/0701216}{{\tt arXiv:hep-th/0701216}}.

\bibitem{Duff:1995sm}
M.~J. Duff, J.~T. Liu, and J.~Rahmfeld, ``{Four-dimensional
  string-string-string triality},''
  \href{http://dx.doi.org/10.1016/0550-3213(95)00555-2}{{\em Nucl. Phys.} {\bf
  B459} (1996)  125--159},
\href{http://arxiv.org/abs/hep-th/9508094}{{\tt arXiv:hep-th/9508094}}.

\bibitem{Behrndt:1996hu}
K.~Behrndt, R.~Kallosh, J.~Rahmfeld, M.~Shmakova, and W.~K. Wong, ``{STU black
  holes and string triality},''
  \href{http://dx.doi.org/10.1103/PhysRevD.54.6293}{{\em Phys. Rev.} {\bf D54}
  (1996)  6293--6301},
\href{http://arxiv.org/abs/hep-th/9608059}{{\tt arXiv:hep-th/9608059}}.

\bibitem{Billo:1999ip}
M.~Billo {\em et al.}, ``{The 0-brane action in a general D = 4 supergravity
  background},'' \href{http://dx.doi.org/10.1088/0264-9381/16/7/313}{{\em
  Class. Quant. Grav.} {\bf 16} (1999)  2335--2358},
\href{http://arxiv.org/abs/hep-th/9902100}{{\tt arXiv:hep-th/9902100}}.

\bibitem{deBoer:2006vg}
J.~de~Boer, M.~C.~N. Cheng, R.~Dijkgraaf, J.~Manschot, and E.~Verlinde, ``{A
  farey tail for attractor black holes},'' {\em JHEP} {\bf 11} (2006)  024,
\href{http://arxiv.org/abs/hep-th/0608059}{{\tt arXiv:hep-th/0608059}}.

\bibitem{Bena:2008wt}
I.~Bena, N.~Bobev, and N.~P. Warner, ``{Spectral Flow, and the Spectrum of
  Multi-Center Solutions},''
\href{http://arxiv.org/abs/0803.1203}{{\tt arXiv:0803.1203 [hep-th]}}.

\bibitem{Moore:1998pn}
G.~W. Moore, ``{Arithmetic and attractors},''
\href{http://arxiv.org/abs/hep-th/9807087}{{\tt arXiv:hep-th/9807087}}.

\bibitem{Denef:2002ru}
F.~Denef, ``{Quantum quivers and Hall/hole halos},'' {\em JHEP} {\bf 10} (2002)
   023,
\href{http://arxiv.org/abs/hep-th/0206072}{{\tt arXiv:hep-th/0206072}}.

\bibitem{Gutowski:2003rg}
J.~B. Gutowski, D.~Martelli, and H.~S. Reall, ``{All supersymmetric solutions
  of minimal supergravity in six dimensions},''
  \href{http://dx.doi.org/10.1088/0264-9381/20/23/008}{{\em Class. Quant.
  Grav.} {\bf 20} (2003)  5049--5078},
\href{http://arxiv.org/abs/hep-th/0306235}{{\tt arXiv:hep-th/0306235}}.

\bibitem{Iizuka:2005uv}
N.~Iizuka and M.~Shigemori, ``{A note on D1-D5-J system and 5D small black
  ring},'' {\em JHEP} {\bf 08} (2005)  100,
\href{http://arxiv.org/abs/hep-th/0506215}{{\tt arXiv:hep-th/0506215}}.

\bibitem{Dabholkar:2005qs}
A.~Dabholkar, N.~Iizuka, A.~Iqubal, and M.~Shigemori, ``{Precision microstate
  counting of small black rings},''
  \href{http://dx.doi.org/10.1103/PhysRevLett.96.071601}{{\em Phys. Rev. Lett.}
  {\bf 96} (2006)  071601},
\href{http://arxiv.org/abs/hep-th/0511120}{{\tt arXiv:hep-th/0511120}}.

\bibitem{Balasubramanian:2005qu}
V.~Balasubramanian, P.~Kraus, and M.~Shigemori, ``{Massless black holes and
  black rings as effective geometries of the D1-D5 system},'' {\em Class.
  Quant. Grav.} {\bf 22} (2005)  4803--4838,
\href{http://arxiv.org/abs/hep-th/0508110}{{\tt arXiv:hep-th/0508110}}.

\bibitem{David:2002wn}
J.~R. David, G.~Mandal, and S.~R. Wadia, ``{Microscopic formulation of black
  holes in string theory},''
  \href{http://dx.doi.org/10.1016/S0370-1573(02)00271-5}{{\em Phys. Rept.} {\bf
  369} (2002)  549--686},
\href{http://arxiv.org/abs/hep-th/0203048}{{\tt arXiv:hep-th/0203048}}.

\bibitem{Sen:1995in}
A.~Sen, ``{Extremal black holes and elementary string states},''
  \href{http://dx.doi.org/10.1142/S0217732395002234}{{\em Mod. Phys. Lett.}
  {\bf A10} (1995)  2081--2094},
\href{http://arxiv.org/abs/hep-th/9504147}{{\tt arXiv:hep-th/9504147}}.

\bibitem{Dabholkar:2006za}
A.~Dabholkar, N.~Iizuka, A.~Iqubal, A.~Sen, and M.~Shigemori, ``{Spinning
  strings as small black rings},'' {\em JHEP} {\bf 04} (2007)  017,
\href{http://arxiv.org/abs/hep-th/0611166}{{\tt arXiv:hep-th/0611166}}.

\bibitem{Skenderis:2006ah}
K.~Skenderis and M.~Taylor, ``{Fuzzball solutions and D1-D5 microstates},''
  \href{http://dx.doi.org/10.1103/PhysRevLett.98.071601}{{\em Phys. Rev. Lett.}
  {\bf 98} (2007)  071601},
\href{http://arxiv.org/abs/hep-th/0609154}{{\tt arXiv:hep-th/0609154}}.

\bibitem{Kanitscheider:2006zf}
I.~Kanitscheider, K.~Skenderis, and M.~Taylor, ``{Holographic anatomy of
  fuzzballs},'' {\em JHEP} {\bf 04} (2007)  023,
\href{http://arxiv.org/abs/hep-th/0611171}{{\tt arXiv:hep-th/0611171}}.

\bibitem{Gaiotto:2004ij}
D.~Gaiotto, A.~Strominger, and X.~Yin, ``{Superconformal black hole quantum
  mechanics},'' {\em JHEP} {\bf 11} (2005)  017,
\href{http://arxiv.org/abs/hep-th/0412322}{{\tt arXiv:hep-th/0412322}}.

\bibitem{Strominger:1996sh}
A.~Strominger and C.~Vafa, ``{Microscopic Origin of the Bekenstein-Hawking
  Entropy},'' \href{http://dx.doi.org/10.1016/0370-2693(96)00345-0}{{\em Phys.
  Lett.} {\bf B379} (1996)  99--104},
\href{http://arxiv.org/abs/hep-th/9601029}{{\tt arXiv:hep-th/9601029}}.

\bibitem{Raeymaekers:2007ga}
J.~Raeymaekers, ``{Near-horizon microstates of the D1-D5-P black hole},''
  \href{http://dx.doi.org/10.1088/1126-6708/2008/02/006}{{\em JHEP} {\bf 02}
  (2008)  006},
\href{http://arxiv.org/abs/0710.4912}{{\tt arXiv:0710.4912 [hep-th]}}.

\bibitem{Maharana:1992my}
J.~Maharana and J.~H. Schwarz, ``{Noncompact symmetries in string theory},''
  \href{http://dx.doi.org/10.1016/0550-3213(93)90387-5}{{\em Nucl. Phys.} {\bf
  B390} (1993)  3--32},
\href{http://arxiv.org/abs/hep-th/9207016}{{\tt arXiv:hep-th/9207016}}.

\bibitem{Sen:1994fa}
A.~Sen, ``{Strong - weak coupling duality in four-dimensional string theory},''
  \href{http://dx.doi.org/10.1142/S0217751X94001497}{{\em Int. J. Mod. Phys.}
  {\bf A9} (1994)  3707--3750},
\href{http://arxiv.org/abs/hep-th/9402002}{{\tt arXiv:hep-th/9402002}}.

\end{thebibliography}\endgroup
\bibliographystyle{utphys}

\end{document}